\documentclass{aa}  
\usepackage{graphicx}
\usepackage{txfonts}
\usepackage{natbib}
\usepackage{ulem} % pour faire sout et uwave
\newcommand{\cha}[1]{#1}

\newcommand{\chc}[1]{{#1}}
\renewcommand{\d} {\mathrm{d}}
\newcommand{\ds}[1]  {\frac{\d{#1}}{\d s}}
\bibpunct{(}{)}{;}{a}{}{,} % to follow the A&A style
\begin{document}
\title{Coronal heating in coupled photosphere-chromosphere-coronal systems:
turbulence and leakage}
\titlerunning {Turbulence and leakage in coronal heating}
%\title {Coupling photosphere and corona: turbulent regime}

   \author{A. Verdini
          \inst{1}
          R. Grappin
          \inst{2,3}
          M. Velli
          \inst{4}
          }
          
   \institute{Solar-Terrestrial Center of Excellence - SIDC, Royal
Observatory of Belgium, Bruxelles \\
              \email{verdini@oma.be}
          \and
             LUTH, Observatoire de Paris, Meudon
          \and
             LPP, Ecole Polytechnique, Palaiseau
          %\and
          %   Dipartimento di Astronomia, Firenze
           \and
             JPL, California Institute of Technology, Pasadena            
           }

   \date{Received ; accepted }

% \abstract{}{}{}{}{} 
% 5 {} token are mandatory
 
 \abstract
 { % context
Coronal loops act as a resonant cavity for low frequency fluctuations that are transmitted from the deeper layers of the solar atmosphere. Such fluctuations are amplified in the corona and lead to the development of turbulence that in turn is able to dissipate the accumulated energy, thus heating the corona.
However trapping is not perfect, some energy leaks down to the chromosphere on a long timescale, thus limiting the turbulent heating.
}
{We consider the combined effects of turbulence and energy leakage from the corona to the photosphere in determining the turbulent energy level and associated heating rate in models of coronal loops which include the chromosphere and transition region.
}
 {We use a piece-wise constant model for the Alfv\'en speed in loops and a Reduced MHD - Shell model to describe the interplay between turbulent dynamics in the direction perpendicular to the mean field and propagation along the field. Turbulence is sustained by incoming fluctuations which are equivalent, in the line-tied case, to forcing by the photospheric shear flows. 
While varying the turbulence strength, we compare systematically the average
coronal energy level and dissipation in three models with increasing
complexity: the classical closed model, the open corona, and the open corona
including chromosphere (or 3-layer model), the latter two models allowing
energy leakage.
}
 {We find that:
 (i) Leakage always plays a role: even at for strong turbulence,
 the dissipation time never becomes much lower than the leakage time, at least in the three-layer
 model. Hence, the
 energy as well as the dissipation levels are systematically lower than in the line-tied model.
(ii) In all models, the energy level is close to the resonant prediction, i.e., assuming effective turbulent correlation time longer than the Alfv\'en coronal crossing time.
(iii) The heating rate is close to the value given the ratio of photospheric energy divided by the Alfv\'en crossing time.
(iv) The coronal spectral range is divided in two, an inertial range with $5/3$ spectral slope, and a large scale peak where nonlinear couplings are inhibited by trapped resonant modes.
(v) In the realistic 3-layer model, the two-component spectrum leads to a
global decrease of damping equal to Kolmogorov damping reduced by a factor
$u_{rms}/V_a^c$ where $V_a^c$ is coronal Alfv\'en speed.
    }
{}


   \maketitle
%
%________________________________________________________________

\section{Introduction}
%In the line-tied approximation, only turbulent dissipation can limit the energy accumulation in the corona.
%In reality, another effect can limit the energy accumulation, namely the leakage through the T.R..
%If the leakage time is comparable or shorter than the dissipation time, then this might might have several consequences: 
%(i) the coronal trapped energy could decrease as the trapping time would be reduced
%(ii) the dissipation (thus the heating rate) could vanish if there is no time for nonlinear coupling to occur before energy returns back to the photosphere
Solving the coronal heating problem involves understanding how fast 
magnetic energy can be accumulated in the corona and how fast this energy is dissipated.
We investigate this problem by considering a model loop in which 
kinetic and magnetic energies are injected into the corona in the form of
Alfv\'en waves generated by photospheric motions.  A large body of work has been 
devoted to this problem, 
 \citep{Milano_al_1997, Dmitruk_al_2003, Rappazzo_al_2007,
Rappazzo_al_2008, Nigro_al_2004, Nigro_al_2005, Nigro_al_2008, Buchlin_Velli_2007}:
we consider here a previously neglected effect which plays a large role in regulating the turbulent energy balance in the corona, namely the leakage of coronal energy back down to the photosphere.

A solar loop can be described as a bundle of magnetic field lines that expand into the corona but are rooted in 
the denser photosphere at two (distant) points, so that their length is typically much greater than the
transverse scale. The magnetic field is therefore mostly along the direction of the loop, and provided 
the transverse magnetic field is not too strong the curvature of the loop may be neglected.
In addition, if the ratio of the plasma to magnetic field pressures is small the motions are predominantly incompressible,
so the transverse structure in density may be neglected compared to the gravitational stratification, while the expansion of the field from the denser layers of the photosphere and chromosphere into the corona may be taken into account via gradients along the field of the Alfv\'en speed.
The resulting, simplified coronal loop retains the basic ingredients which lead to heating: turbulent coupling and propagation through a stratified atmosphere where stratification appears as an increase of the Alfv\'en speed from photosphere to corona.

The  stratification is characterized by the ratio of mean Alfv\'en speeds  in the photosphere ($V_a^0$) and in the corona ($V_a^c$) which is a small parameter:
\begin{equation}
\epsilon=V_a^0/V_a^c<<1
\label{epsilon}
\end{equation}
The part of the wave spectrum incoming into the corona that we shall consider here is the low frequency part, for which the Alfv\'en speed contrast is seen by waves of frequency $\omega$ as a sharp transition.
This occurs if
\begin{equation}
 \omega\lesssim\mathrm{max}(|{\bf \nabla V_a}|)\approx
(V_a^c-V_a^0)/H\approx5-10~\mathrm{Hz}
\end{equation}
for  $V_a^c\approx 2000~\mathrm{km/s}$ and a transition region thickness of about $H=200~\mathrm{km}$.
For these low frequencies, the transition region (T.R.) acts as a transmitting and reflecting barrier, with the important property that the transmission is not symmetric, so that a coronal loop acts
as a cavity which resonates at specific frequencies, based on the Alfv\'en
crossing time $t_a^c=L_c/V_a^c$ \chc{($L_c$ is the length of the coronal
part of loop)}:
\begin{equation}
\omega = n\pi V_a^c/L_c=n\pi/t_a^c
\end{equation}
with $n=0,1...$ \citep{Ionson_1982,Hollweg_1984a}.

The cavity is perfectly insulated in the limit of infinite Alfv\'en speed
contrast, i.e. $\epsilon = 0$, which corresponds to the so-called line-tied
limit. In this limit, the corona exerts no feedback on the solar surface. The
zero frequency resonance is clearly distinct from the finite frequency
resonances; in the former, the coronal magnetic energy grows without bounds
while the kinetic energy remains finite \citep{Parker_1972,Rappazzo_al_2007}; in the latter case, both magnetic and kinetic coronal energies grow at equipartition.

In reality, the trapped energy is limited, because the cavity looses energy by
two different mechanisms: damping (turbulent or not), and leakage, due to the finite Alfv\'en
speed contrast. The leakage time is given by \citep{Hollweg_1984a, Ofman_2002,
Grappin_al_2008}:
\begin{equation}
t_L = L_c/V_a^0
\label{fuite}
\end{equation}
Note that the leakage time is much greater than the Alfv\'en crossing time, since $t_a^c = \epsilon t_L$
\footnote{\chc{As will be seen in sec.~\ref{sec:3lay}, eq.~\ref{fluxOK}, at every reflection a fraction
$\epsilon$ of the coronal energy leakes from the transition region down to the
chromosphere, so
one needs $1/\epsilon$ reflections to evacuate the coronal energy, i.e. a
timescale $t_a^c/\epsilon$.}}.
The dissipation rate of the loop will thus depend on
(i) the energy input into the corona, as well as its frequency distribution (resonant or not)
(ii) what part of the energy input goes into heat what part returns back to the solar surface (leakage).

In the previous works starting with \cite{Hollweg_1984a}, 
it has always been assumed that the leakage time was long compared to the (turbulent) dissipation time, thus leakage was neglected (line-tied limit).
% andrea
Because neglecting leakage implies neglecting the back-reaction of the corona on the 
deeper layers, in the line-tied limit the velocity can be imposed at the coronal base. 
This is justified if the leakage time is larger than the coronal dissipation time.
Estimating the latter to be given by the photospheric turnover time $t_{NL}^0=l_\bot/(2\pi U_0)$, we have for the ratio of the two time scales:
\begin{eqnarray}
\chi_L & = & t_L/t_{NL}^0 \simeq (L/l_\bot) 2\pi (U_0/V_a^0) \gtrsim 2 \pi
\end{eqnarray}
Since the coronal energy per unit mass is expected to 
reach larger values than at the surface, this largely justifies 
neglecting leakage. However, identifying the dissipation time with the turnover
time might be erroneous, as turbulence, at least in some simulations (e.g.,
\citealt{Nigro_al_2008}) shows a high degree of intermittency, so that the dissipation time is orders of magnitude larger than such simple estimates. 

\begin{figure}[t]
\begin{center}
\includegraphics [width=\linewidth]{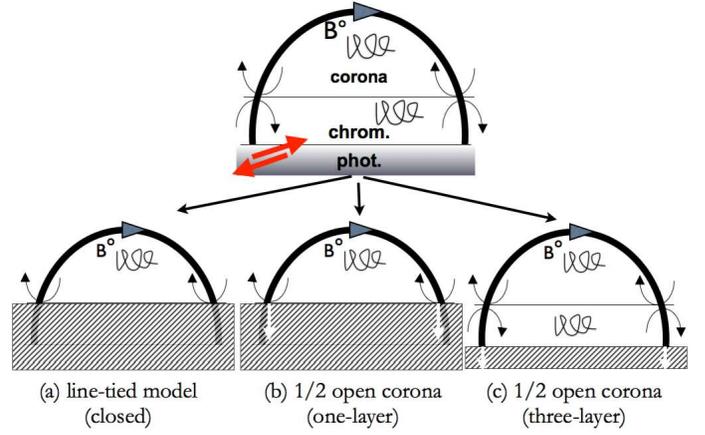}
\caption{
Sketch of the coronal heating process.
Above : the general problem of photospheric injection, transmission, turbulent
dissipation, and leakage back to the photosphere. Red thick arrows at the left foot point representing the surface shear forcing.
Below: the three numerical models considered in this paper:
(a) closed model (no leakage) with imposed velocity at the coronal base
(b) semi-transparent corona with imposed wave input at the coronal base
(c) semi-transparent corona including chromospheric turbulence, with imposed wave input at the chromospheric base.
Thin arrows indicate the wave reflection and transmission, white thick arrows represent the leakage out of the numerical domain.
}
\label{figmodOK}
\end{center}
\end{figure}

This motivates us to relax the line-tied hypothesis, using models of turbulent loops that include leakage. 
The problem becomes then more complex, as the velocity boundary conditions are no longer fixed, the velocity being the sum of the incoming coronal base field and the outcoming coronal signal.
We will consider two versions of the problem that includes leakage. In the first version,
which will be called the one-layer model, we simply change the boundary
conditions at the coronal base, taking leakage into account. The incoming spectrum depends partly of the (given)
signal assumed given by the chromospheric layers below, partly on the signal propagating downward from the corona and being largely (but not fully) reflected.
In the second version, which will be called the three-layer model, the domain is enlarged to 
include two chromospheric layers.
In that case, the signal propagating upward from the coronal base is still more uncontrolled than in the previous case, as the chromospheric turbulence which develops and determines the state of the coronal base is not directly predictable from the photospheric input.
Fig.~\ref{figmodOK} summarizes the models: the classical closed model, and the two versions including leakage.

To describe the turbulence dynamics along the loop, we will use the Shell model
for Reduced MHD \citep{Nigro_al_2005, Buchlin_Velli_2007}.
\chc{Shell models of turbulence share with full turbulence power-law energy
spectra, as well as chaotic (intermittency) properties which are very close to
direct numerical simulations of primitive MHD equations
\citep{Gloaguen_al_1985,Biskamp_1994}}
The system will be forced by introducing DC fluctuations, i.e., a spectrum of fluctuations at different perpendicular scales that is constant in time.

We will show that the finite leakage time leads to significant differences with previous results obtained using line-tied boundary conditions. The plan is the following. The next section deals with basic physics, model equations and parameters. Section three deals with simple phenomenology. Results are in section four, section five contains the discussion.

%%%%%%%%%%%%%%
\section{Basic physics, model equations and parameters}
%{\bf deleted in MV\\
%In what follows, we define first the atmosphere model which determines  the wave boundary  (jump) conditions conditions at transition region (T.R.) interface, then the turbulence model, then how we implement the jump conditions in the model (for the two semi-transparent variants), and finally the parameters.
%The jump conditions in two different ways: either we will restrict the numerical domain to contain the sole corona (so that the boundaries coincide with the transition region), and in that case we will use the jump conditions as boundary conditions, or we will include the chromosphere within the numerical domain, and in that case the jump conditions are used at each time step to generate the new transmitted and reflected signals through the transition region. 
%}
%
\subsection{Three-layer atmosphere: linear reflection/transmission
laws}\label{sec:linear}
\begin{figure}[t]
\begin{center}
\includegraphics [width=\linewidth]{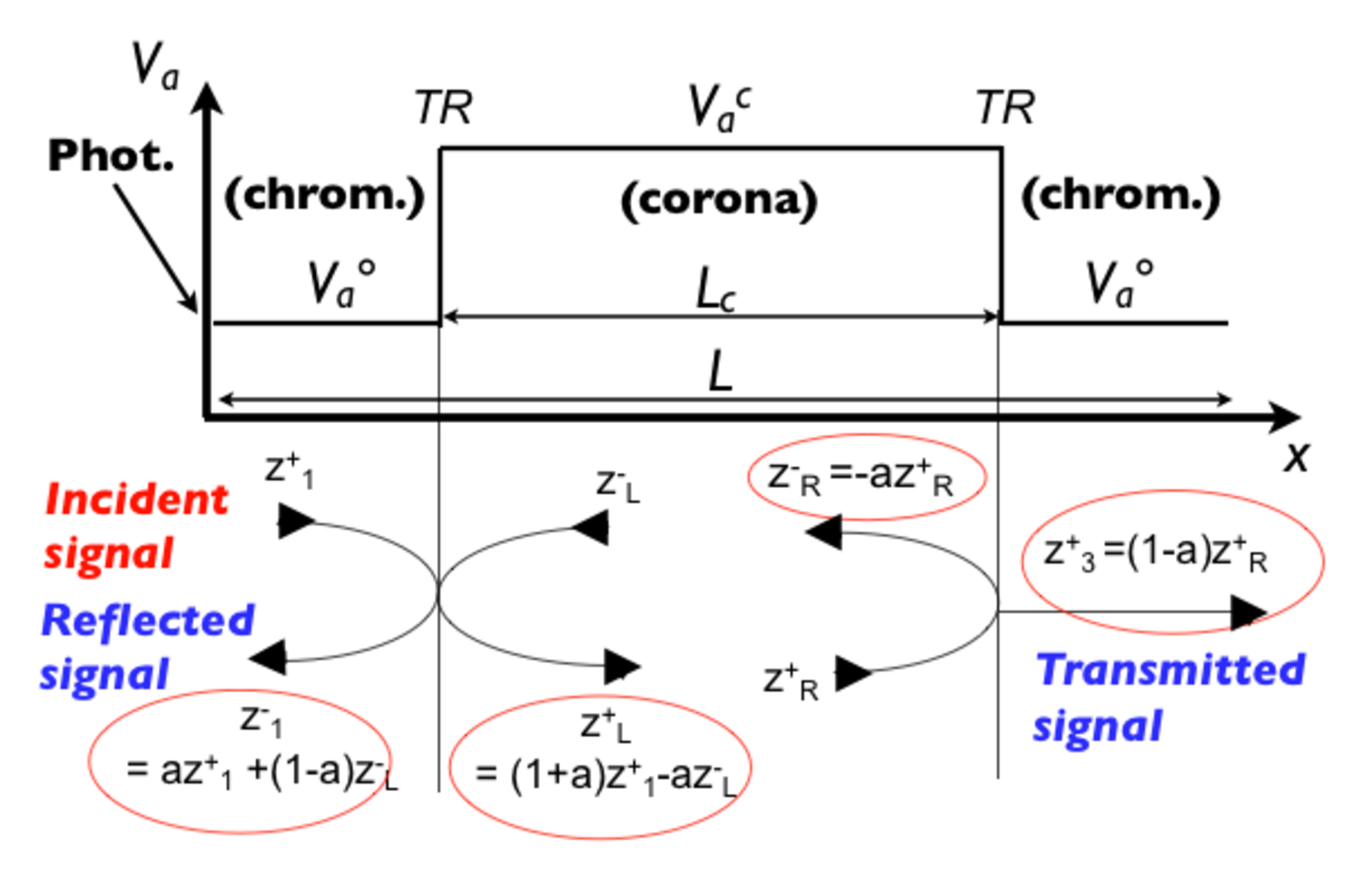}
\caption{
The three-layer model: sketch of the transmission and reflection properties of transverse fluctuations at the coronal bases of a magnetic loop with piece-wise constant Alfv\'en speed, in the particular case considered here (no input from right chromosphere). 
}
\label{fig1}
\end{center}
\end{figure}

We begin by describing our model atmosphere and the properties of linear Alfv\'en wave propagation within such an atmosphere. The atmosphere is considered to be stratified in the vertical direction, with three successive layers representing a left photosphere/chromosphere, the corona, and a right photosphere/chromosphere.  The atmosphere is threaded by a vertical uniform field $B_0$ along which Alfv\'en waves propagate.  In each of these three layers, the Alfv\'en speed is constant, so that a progressive Alfv\'en wave propagates at constant speed without deformation.  When a wave encounters a density jump interface, the velocity and magnetic field fluctuations, which are parallel to
the interface, are continuous. The proper Alfv\'en modes propagating in opposite directions along the loop are defined by the Els\"{a}sser variables:
\begin{equation}
z^\pm = u \mp b/\sqrt \rho
\label{elsaesser}
\end{equation}
where $\rho$ is the density and $ u, b$ are the velocity and magnetic field fluctuations, which are in planes parallel to the photosphere/corona transition region.
Assuming a positive mean field $B_0$, the quantity $z^+$ will propagate to the right and the quantity $z^-$ to the left.
It is immediately seen from this definition that the density jump at the transition region will determine a wave amplitude jump  of order $1/\sqrt \rho=1/\epsilon$.
The derivation of the jump relations may be found in \citep{Hollweg_1984a}. Continuity of the velocity and magnetic field fluctuations at the two interfaces imply the following relations between wave amplitudes respectively at left and right boundaries:
\begin{eqnarray}
z^+_1 + z^-_1 = z^+_L +z^-_L, \ z^+_1 - z^-_1 = (z^+_L - z^-_L) \ \epsilon \nonumber \\
z^+_3 + z^-_3 = z^+_R +z^-_R, \ z^+_3 - z^-_3 = (z^+_R - z^-_R) \ \epsilon
\label{continuity}
\end{eqnarray}
We use $1,L$ to denote the amplitudes at the left T.R. (resp. $1$ on the
photospheric side, $L$ on the coronal side), and $3,R$ to denote the amplitudes
at the right T.R. (resp. $R$ on the coronal side, $3$ on the photospheric side); see Fig.~\ref{fig1}:
\begin{eqnarray}
z^\pm_1=z^\pm(x=0^-) \\
z^\pm_L=z^\pm(x=0^+) \\
z^\pm_R=z^\pm(x=L_c^-) \\
z^\pm_3=z^\pm(x=L_c^+) 
\end{eqnarray}
\chc{with the exponents $+$ or $-$ in $0$ and $L_c$ indicating whether we are on the right or the left side of the
two transition regions, located respectively at $x=0$ and $x=L_c$.}

To obtain the jump conditions to be effectively implemented in the three-layer
model, we rewrite Eqs.~\ref{continuity} as follows. We denote by input what
goes into the corona and output what goes out. The coronal inputs $z^+_L$ and
$z^-_R$ are expressed in terms of the chromospheric inputs ($z^+_1$ and
$z^-_3$) and the coronal outputs ($z^-_L$ and $z^+_R$). Similarly
the reflected chromospheric signals $z^-_1$ and $z^+_3$ are expressed in terms of the chromospheric inputs and of the coronal outputs:
\begin{eqnarray}
z^+_L= (1+a)z^+_1 -a z^-_L \nonumber \\
z^-_R = (1+a)z^-_3 -a z^+_R \nonumber \\
z^-_1= (1-a)z^-_L + a z^+_1 \nonumber \\
z^+_3= (1-a)z^+_R + a z^-_3
\label{transmission}
\end{eqnarray}
The parameter $a$:
\begin{equation}
a= (1-\epsilon)/(1+\epsilon)
\label{a}
\end{equation}
is the reflection coefficient.
It is instructive to consider the limit $\epsilon = 0$. Then the coronal reflection coefficient $a$ becomes unity. 
In this case, the velocity at the left coronal boundary is exactly $z^+_1$, that is, specifying the chromospheric 
input is the same as specifying the velocity (and the same at the right coronal boundary): this is the well-known
line-tied limit. In this limit, the magnetic field fluctuation is not specified and depends on the coronal evolution, since one has: $b_L/\sqrt \rho = -z^+_1 + z^-_L$.
Returning to the general case with a non-zero Alfv\'en speed ratio $\epsilon$, we see that specifying the chromospheric input does not directly determine the velocity at the T.R. either.  We will choose here to consider a non-zero input only from the left foot point (boundary), in order to better follow the propagation of the incident signal. 

In the early work by \citet{Hollweg_1984a},
the three-layer model was studied analytically, with a damping term
representing the effects of turbulence. 
%\uwave{The jump conditions at the T.R. have been recently employed in numerical
%simulations that combine the turbulent dynamics to the propagation properties
%of Alfv\'en waves \citep{Verdini_al_2009, Verdini_al_2010,
%VanBallegooijen_al_2011}}.
As we said, turbulent dissipation is highly intermittent thus requiring a
description that goes beyond a simple damping term.
We now define the nonlinear part of the model, i.e., the
turbulence model.

\chc{
The jump conditions just described are not specific of a linear framework.
In the general case where the waves have a perpendicular structure and interact nonlinearly,
the jump conditions hold as well.
In the final model to be detailed now, where the wave amplitudes depend on the coordinate along the loop and on an index $n$ representing the perpendicular wavenumber $k_n$, the jump conditions
are valid for each Fourier coefficient $z^\pm_n(x)=z^\pm(x,k_n)$ at $x=0$ and $x=L_c$, if
$0$ and $L_c$ are the two coordinates of the transition region.
In the following the integer subscripts $1$ and $3$ will be used to label the
layers as in Fig.~\ref{fig1}, while the
fourier modes will be labelled with the generic index $n$}.

\subsection{Nonlinear model: Shell model for Reduced MHD}
In addition to the linear propagation of perturbations parallel to the loop mean field,  
we consider the waves to have perpendicular structure, so that the wave-vectors also have non-vanishing
components in planes perpendicular to the mean magnetic field. In this transverse direction nonlinear interactions
between different perpendicular modes occur, while the dynamics of the parallel propagation (for a given perpendicular mode) remains purely linear. This model, known as Reduced MHD or RMHD \citep{Strauss_1976}, is believed to be well adapted to situations with a large uniform axial field $B_0$ compared to perturbation amplitudes and strong anisotropy in the sense
that the scales perpendicular to the field are shorter than the length of the coronal loop \citep{Rappazzo_al_2007}:
\begin{equation}
{{\partial \vec z^\pm_{\perp}} \over {\partial t}}\mp \frac{B_0}{\sqrt{\rho}} { {\partial \vec z^\pm_{\perp}} \over {\partial x} }
=-(\vec {z^\mp_{\perp}} \cdot \vec \nabla \vec z^\pm_{\perp})
- {1\over \rho} \vec \nabla _{\perp} (p^T)
+\nu \nabla ^2 _{\perp} \vec z^\pm_{\perp},
\label{s1}
\end{equation}
where we have taken identical kinematic viscosity and resistivity, the density is uniform in the direction orthogonal to the field and the total pressure gradient guarantees the incompressibility of the $\vec z^\pm$ fields via the Poisson equation
\begin{equation}
\nabla^2 _{\perp} (p^T) = - \vec \nabla\cdot (\vec {z^\mp_{\perp}} \cdot \vec \nabla \vec z^\pm_{\perp}).
\end{equation}

A second approximation consists in transforming the perpendicular nonlinear couplings by
replacing them, at each point of the $x$ coordinate mesh along the mean
field direction, by a dynamical system defined in Fourier space, which allows
reaching a very high Reynolds number compared to genuine Reduced MHD. This is
known as the Shell model for RMHD or hybrid Shell model \citep{Nigro_al_2005, Buchlin_Velli_2007}.
\chc{The Reynolds number gain can be quantified as follows. 
Assume $K$ is the perpendicular resolution (ratio from largest to smallest scales.
Assume also the parallel resolution scales as $K^{2/3}$. 
When passing from the RMHD to Shell RMHD the number of degrees of freedom 
changes  from $K^{2+2/3}$ to $K^{2/3} Log_2(K) \simeq K^{2/3}$ (see below).
The CPU time required to describe the same large scale evolution is proportional to this number multiplied by $K$. 
Conversely, the resolution reachable goes as the CPU time $T$ as $T^{3/11}$ in
the RMHD case and as $T^{3/5}$ in the Shell RMHD case, thus passing from a
resolution $K_0$ to a resolution $K_0^{11/5}$. The same is true for the
Reynolds number (which goes as a power of the resolution $K$), hence typically
passing from $10^3$ to $10^6$.
}

Coronal heating driven by photospheric motions has been studied using both RMHD and RMHD Shell models in a one-layer atmosphere (corona) version, with uniform Alfv\'en speed and closed (line-tied) boundaries, i.e. imposing the photospheric perpendicular velocity at loop foot points. Here we will use an RMHD-Shell model, but in the three-layer context, that is, including the linear jump laws defined previously at the transition region for each of the perpendicular wave number.

The Shell model is characterized by the number $N+1$ of perpendicular wave modes, each being characterized by a perpendicular wave number, with amplitudes $z^\pm_n$ (the direction of the wave vector is not specified in the model), with the following discretization:
\begin{equation}
k_n = 2^n k_0 \\ \\ n=0... N
\end{equation}
Starting from the RMHD equations, one can write the following simplified equations (see \citealt{Buchlin_Velli_2007} for the full equations with non
homogeneous density):
\begin{eqnarray}
\partial_t z^+_n + V_a\partial_x z^+_n = T_n^+ - \nu k_n^2 z^+_n \nonumber \\
\partial_t z^-_n - V_a\partial_x z^-_n = T_n^- - \nu k_n^2 z^-_n
\label{basic}
\end{eqnarray}
where $V_a$ is either $V_a^0$ (chromosphere) or $V_a^c$ (corona), $\nu$ is the kinematic viscosity (equal to the magnetic diffusivity), and the $T^\pm_n$ are the nonlinear terms which are a sum of terms of the form: $T^\pm_n = A k_{m}z^\mp_p z^\pm_q$ with $m$, $p$, $q$ being close to $n$
(see \citealt{Biskamp_1994, Giuliani_Carbone_1998} for the full expression of
$T^\pm_n$).

From the basic Eqs.~\ref{basic}, one can deduce the (exact) energy budget equation of a flux tube of length $L$, section $\pi l_{\bot 0}^2$ and density $\rho$, assumed constant, as:
\begin{equation}
\d E/\d t = F(t) - D(t)
\label{balance}
\end{equation}
$E$ is the total energy, $F$ the energy flux and $D$ the energy dissipation
rate defined as:
\begin{eqnarray}
E&=& M \frac{1}{2L} \int_0^L \d x \ (u^2+b^2/\rho) \nonumber \\ 
&=& M \frac{1}{4L} \int_0^L \d x \;
\left[(z^+)^2+(z^-)^2\right]  
\label{fluxa}\\
F&=& M V_a \frac{1}{4L}  \left[(z^+_0)^2 - (z^+_L)^2 +
(z^-_L)^2 - (z^-_0)^2\right] 
\label{flux} \\
D&=&M \frac{1}{2L}  \int_0^L \d x \; \sum_{n=0}^N \nu k_n^2 (z_n^{+2}+z_n^{-2}) 
\label{fluxc}
\end{eqnarray}
Here $M=\pi l_{\bot 0}^2 L \rho$ is the mass of the loop system, $u^2$ and
$b^2/\rho$, $(z^+)^2$, $(z^-)^2$ are the sum respectively of the
energies per unit mass in all the modes $n=0..N$.
When applying Eq.~\ref{balance}
to the corona, we take 
$\rho_c=\epsilon^2 \rho_0$, $V_a=V_a^c$ and the subscripts (integration
interval) $0,~L$ represent the left and right coronal
boundaries respectively (not including the chromosphere when it is present). 
Note that the parameter $l_{\bot 0}$ stands for the largest scale available in the simulation, which is always in all runs 
$l_{\bot 0} = 4 l_{\bot}$.
In the following, we will use the notations $E, D, F$ as defined in Eqs.~\ref{flux}-\ref{fluxc} but always normalized by the total mass $M$ of the loop system, so obtaining average energies and dissipation rates per unit mass.

Several remarks are in order.
First,  the nonlinear terms don't appear in the energy budget equation~\ref{balance}, because the total energy is conserved by nonlinear coupling, as well in the Reduced MHD equations as in the presently used Shell model version of the equations. 
Second the energy accumulated or lost by the corona is not directly controlled. 
Indeed, the energy flux entering the corona (Eq.~\ref{flux}, see also the more explicit Eq.~\ref{fluxOK} below) is determined by the difference between the incoming and outcoming energies at the two transition regions; 
as will be made clear in the next subsection, the boundary conditions fix the incoming amplitudes, possibly in terms of the outcoming amplitudes, but not the energies.

\subsection{Boundary and jump conditions for three- and one-layer model}
As a rule, boundary conditions are defined by imposing the value of $z^+_n$
at $x=0$ (the rightward propagating wave amplitude) and the value of $z^-_n$ at the boundary $x=L$ (the leftward wave amplitude).
The jump conditions at the transition region

\subsubsection{Closed model (line-tied)}

The loop contains only the corona.
The usual closed or line-tied model has for boundary conditions:
\begin{eqnarray}
z^+_n(x=0,t) = 2U^0_n(t) - z^-_n(x=0,t) \nonumber \\
z^-_n(x=L,t) = 2U^L_n(t) - z^+_n(x=L,t) 
\label{closed}
\end{eqnarray}
The previous equation results from Eqs.~\ref{transmission} with $a=1$, $z^+_L
\equiv z^+_n(x=0,t)$, and $z^-_R \equiv z^-_n(x=L,t)$.
The $z^+_L$ ($z^-_R$) signal in the corona is obtained by prescribing the \textsl{velocity
amplitude} $U^0_n$ ($U^L_n$) of each mode $n$ at the boundary $x=0$ ($x=L$).
\cha{One checks from Eq.~\ref{balance} that when $u_{0,n}=0$, then the energy flux (Eq.~\ref{fluxc}) injected in the domain becomes indeed zero.}

\subsubsection{One-layer model}
In this first model including leakage, the chromosphere is excluded from the domain, 
the domain boundaries coinciding with the T.R., as in the closed model.
The boundary conditions now take the wave jump conditions ($a<1$ in Eqs.~\ref{transmission}) explicitly into account:
\begin{eqnarray}
z^+_n(x=0,t) = (1+a)Z^+_n(t) - az^-_n(x=0,t) \nonumber \\
z^-_n(x=L,t) = (1+a)Z^-_n(t) - az^+_n(x=L,t) 
\label{onelayer}
\end{eqnarray}
The quantities $Z^+_n$ and $Z^-_n$ now denote the prescribed
\textsl{wave amplitudes} entering from the chromospheric side of the transition
region. With reference to Fig.~\ref{fig1} we have $Z^+_n\equiv z^+_1$ and
$Z^-_n \equiv z^-_3$.

\subsubsection{Three-layer model}\label{sec:3lay}
In this second model allowing leakage, the chromosphere is really included within
the domain; in that case the boundary conditions (at the photosphere) are
chosen to be purely open ($a=0$ in Eqs.~\ref{transmission} or
equivalently in Eq.~\ref{onelayer}):
\begin{eqnarray}
z^+_n(x=0,t) = Z^+_n(t) \nonumber \\
z^-_n(x=L,t) = Z^-_n(t)
\label{open}
\end{eqnarray}
The boundaries are open in the sense that incoming waves are defined independently of outgoing waves, which in turn generate no incoming wave, so that they escape freely from the domain: perturbations coming from the loop reach the boundary  and disappear below the boundary without reflection.
Wave reflections and transmissions continuously occur within the domain at the location of the transition regions: there we apply the jump conditions
(Eqs.~\ref{transmission}), for each perpendicular mode $n$.

The three-layer model and the one-layer model with partially reflecting
boundaries are parametrized by the same number $\epsilon$, the
photospheric/coronal Alfv\'en speed ratio. The two models thus both include the
transmission and reflection of waves by the transition region,
but have an important difference.
In the one-layer model, the chromospheric input is specified, as the T.R. coincides with the boundary of the domain.
Instead, in the three-layer model, the chromospheric input ($z^+_1,~z^-_3$
\chc{in Fig~\ref{fig1}}) is not prescribed, since the (prescribed) photospheric input has been modified by turbulence during its propagation through the chromosphere. 
Both models have specific advantages: the three-layer model has more internal degrees of freedom, as it shows two distinct (but coupled) turbulent layers, one in the chromosphere, the other one in the corona; on the other hand, the one-layer model is more directly comparable to the closed line-tied model: the domain is the same (the corona), only the boundary conditions change. We will study both models, with some emphasis on the three-layer model.

In the simulations we present, forcing is applied by injecting upward propagating waves only at the left loop foot point; more
precisely, the leftward propagating amplitude at the right photospheric foot
point $Z^-_n(t)$ will be maintained zero in Eq.~\ref{onelayer}-\ref{open}.
In the particular case of the closed model (Eq.~\ref{closed}), this means that the velocity at the right foot point $U^L_n(t)$ was kept zero.
In this case, it interesting to write down the expression for the net coronal energy flux :
\begin{eqnarray}
F \propto \Sigma_n \bigg \{ (1+a)^2 |z^+_1|^2-2a(1+a)
\mathrm{Re} (z^+_1 \cdot {z^-_L}^\star ) \nonumber \\
- (1-a^2) |z^-_L|^2 -(1-a^2)|z^+_R|^2 \bigg\}
\label{fluxOK}
\end{eqnarray}
In the previous formula, 
\chc{the $\star$ denotes the complex conjugate,}
indices $n$ are assumed for each variable; we have used the 
notations of Fig.~\ref{fig1}, in order for the formula to apply to the three models.
We consider in turn the different terms in the right-hand side. 
The last two terms are always negative: they thus represent a pure leakage (and
they indeed vanish for $a=1$, in the closed or line-tied model).
 The first term is always positive and represents the continuous energy injection.
The second term is fluctuating and is the only term that can cause leakage in the closed model.
Note however that, in the closed case, it is non zero only for the injected modes (which are at large scales, see next subsection), due to the presence of the $z^+_1$ factor: this strongly limits the leakage in the closed case.

%%%%%%%%%%%%%%%%%%%%
\subsection{Parameters and timescales}
The parameters of the model are the length of the chromospheric and coronal part of the loop $L_{ch},~L_c$ respectively, the photospheric-chromospheric Alfv\'en speed, $V_A^0$, the 
Alfv\'en speed contrast $\epsilon$, the width of the loop $l_{\bot0}$, the
turbulent correlation scale $l_\bot$, and the amplitude of the
forcing at the left photosphere, $U_0$.
%(or corona in the closed and 1-layer models). 
In all the models we will always force by injecting an Alfv\'en wave: $U_0$ is the wave amplitude which, generally, is not directly
related to the photospheric velocity shear. Only in the closed model the two
quantity coincide (see section~\ref{sec:linear}).
The input photospheric spectrum will be distributed on the perpendicular
scales $l_\bot$, $l_\bot/2$, $l_\bot/4$, and will have a correlation time given
by $T_f$, which completes the set of parameters.\\
For all the simulation we have set $V_A^0=700~\mathrm{m/s}$,
$L_{ch}=2~\mathrm{Mm}$ (so that $L$ scales with $L_c$ only), i.e. we
assume that photospheric values are independent of the loop length and that
all loops have a transition region. 
We will also set $l_\bot=l_{\bot0}/4$ and $T_f=\infty$.
The rest of the parameters $l_\bot,~L_c,~U_0,~\epsilon$ define
the following physical time scales (i.e., input of the model):
the leakage time, the coronal Alfv\'en time, and the input nonlinear
time (which rules the strength of the turbulence resulting from the
driving):
%\footnote{In the definition of the input nonlinear time, $U_0$ stands for the amplitude of the imposed field at the left boundary. This field correspond to the velocity in the closed model, but for the open models
%(1-layer and 3-layer) it corresponds to the rightward propagating wave,
%$z^+_0=2U_0$. We will use the same notation despite this small difference.}
%, and the correlation time of the forcing:
%{\bf j'ai effac\'e $T_f$ comme il est fix\'e \`a l'infinie et je le dit juste
%avant} 
\begin{eqnarray}
t_L &=& L_c/V_a^0 \\
t_a^c &=& L_c/V_a^c=\epsilon t_L  \\
%t_{NL}^0 &=& l_\bot /(2\pi U_0) \;\;\mbox{ ($U_0$ is the iforced velocity)}\\
t_{NL}^0 &=& l_\bot /(2\pi U_0)
%T_f &=& \infly\;\;\mbox{(forcing correlation time or forcing periodicity)}
\end{eqnarray}
In this work we will also fix $L_c=6~\mathrm{Mm}$ and $\epsilon\approx0.02$,
thus only $t_{NL}^0$ will be varied at fixed $t_a^c$ and $t_L$, by changing the parameters $U_0$ and $l_\bot$ (in a following paper we will study
the effects of varying the leakage and the Alfv\'en time scales).
%, so that
%the loop aspect ratio $l_0/L_c$ is controlled through the ratio $l_\bot/L_c$.
%We will vary the loop aspect ratio $l_\bot/L_c$ (varying $\l_bot$ or $L_c$),
%the input nonlinear timescale (varying $\l_bot$ or $U_0$), and the Alfv\'en
%speed contrast $\epsilon$, which lead to the variation of the 
%two dimensionless time ratios:}
From these characteristic times we define the following dimensionless
parameters which measure the nonlinear term vs the two main linear effects, the
Alfv\'en wave propagation and the leakage:
\begin{eqnarray}
\chi_L & = & t_L/t_{NL}^0   \label{chi_L} \\
\chi_0 & = & t_a^c/t_{NL}^0=\epsilon\chi_L \label{chi_0}
\end{eqnarray}
The parameter $\chi_0$ has been used by \citet{Dmitruk_al_2003,Rappazzo_al_2008, Nigro_al_2008}
to quantify the turbulent behavior in their studies of turbulence forcing
with closed boundaries (corresponding to $\chi_L=\infty$).
\begin{figure}[t]
\begin{center}
\includegraphics [width=\linewidth]{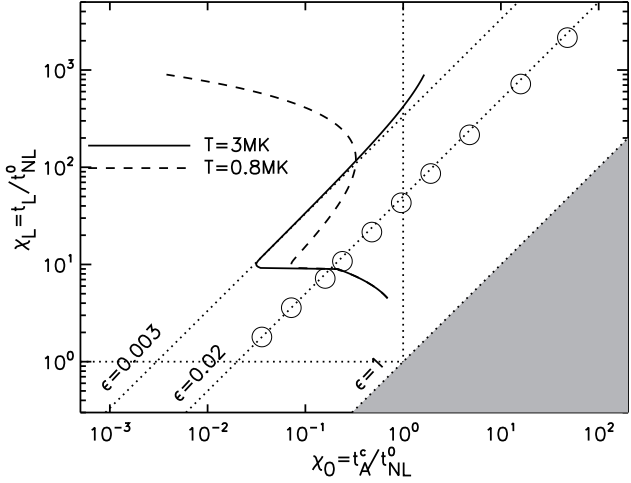}
\caption{
Characteristics of typical solar loops compared with simulation parameters: $\chi_L$ versus $\chi_0$. 
The turnover time is fixed to $t_{NL}^0=1000$~s
and we assume a two-temperature hydrostatic loop model (see appendix). The
black solid and dashed lines are for the two
coronal temperatures of 3~MK and 0.8~MK respectively. 
The circles are for the 3-layer runs $A-I$ (see Table~\ref{table1}).
The diagonal dotted lines are the curve $\chi_L=(1/\epsilon) \chi_0$ for three values of $\epsilon$.
}
\label{fig1a}
\end{center}
\end{figure}

Fig.~\ref{fig1a} shows the plane with $\chi_0$ in abscissa and $\chi_L$ in ordinate.
This plane is divided in four quadrants by the lines $\chi_0=1$ and $\chi_L=1$.
There are actually only three subsets left, as only the subset with $\epsilon < 1$, 
visible as the non shaded region of Fig.~\ref{fig1a}, is permitted, due to the stratification.
Turbulence will be said weak (in the left part) or strong (right part), depending on 
$\chi_0$ being smaller or larger than unity.
In the two upper quadrants, which occupy most of the domain, leakage should be negligible.
Only in the small (left) bottom region should leakage dominate over turbulent loss.

The two curves (solid and dashed) represent each a family of coronal loops of varying length $L$,
build from a two-temperature hydrostatic model (see appendix~\ref{appendix1})
which leads to a function $V_a^c(L)$.
The solid (resp. dashed) curve corresponds to a coronal temperature of $3$ (resp. $0.8$) MK,
and the loop length $L$ increases from bottom to top (i.e., with increasing $\chi_L$) from $3$ to $700$~Mm.
As shown by dotted lines, most of the hot loops show an Alfv\'en speed contrast of $\epsilon = 0.003$, about ten times smaller than the $\epsilon$ value chosen for the 3-layer simulations (see circles).
The choice of relatively large $\epsilon$ values for the simulations comes from the requirement of having a reasonable value for the ratio of integration time to single time step.
%%%%%%%%
% table de simulations
\begin{table}[t]
\centering
\caption{
Parameters for the the simulations}
\begin{tabular}{ccccccc}
\hline
run & $\epsilon$ & $L_c$ & $l_\bot$ & $U_0$ & $\chi_0$ & $\chi_L$ \\
    &  (adim)    &   (Mm)   & (Mm)  & (km/s)& (adim.)  & (adim)  \\
\hline\hline
A  & 0.020 &   6 &  1.500 &    0.05 &      0.04 &      1.8 \\
B  & 0.020 &   6 &  1.500 &    0.10 &      0.07 &      3.6 \\
C  & 0.022 &   6 &  1.500 &    0.20 &      0.16 &      7.2 \\
D  & 0.022 &   6 &  0.500 &    0.10 &      0.24 &     11   \\
E  & 0.022 &   6 &  0.500 &    0.20 &      0.48 &     22   \\
F  & 0.022 &   6 &  0.250 &    0.20 &      0.96 &     43   \\
G  & 0.022 &   6 &  0.125 &    0.20 &      1.9  &     86   \\
H  & 0.022 &   6 &  0.025 &    0.10 &      4.8  &    215   \\
I  & 0.022 &   6 &  0.008 &    0.10 &     16    &    718   \\
L  & 0.022 &   6 &  0.003 &    0.10 &     48    &   2154   \\
\hline\hline
$\mathrm{A_{1L}}$ & 0.020 &   6 &  1.500 &    0.025 &      0.02 &      0.9 \\
$\mathrm{B_{1L}}$ & 0.020 &   6 &  1.500 &    0.10  &      0.07 &      3.6 \\
$\mathrm{C_{1L}}$ & 0.020 &   6 &  1.500 &    0.20  &      0.14 &      7.2 \\
$\mathrm{D_{1L}}$ & 0.020 &   6 &  0.500 &    0.10  &      0.22 &     11   \\
$\mathrm{E_{1L}}$ & 0.020 &   6 &  0.125 &    0.10  &      0.86 &     43   \\
$\mathrm{F_{1L}}$ & 0.020 &   6 &  0.063 &    0.10  &      1.7  &     86   \\
$\mathrm{G_{1L}}$ & 0.020 &   6 &  0.025 &    0.10  &      4.3  &    215   \\
$\mathrm{H_{1L}}$ & 0.020 &   6 &  0.008 &    0.10  &     14    &    718   \\
$\mathrm{I_{1L}}$ & 0.020 &   6 &  0.003 &    0.10  &     43    &   2154   \\
\hline\hline
$\mathrm{A_{cl}}$ &  0 &   6 &  1.500 &    0.025 &      0.02 &  $\infty$ \\
$\mathrm{B_{cl}}$ &  0 &   6 &  1.500 &    0.10  &      0.07 &  $\infty$ \\
$\mathrm{D_{cl}}$ &  0 &   6 &  0.500 &    0.10  &      0.22 &  $\infty$ \\
$\mathrm{E_{cl}}$ &  0 &   6 &  0.125 &    0.10  &      0.86 &  $\infty$ \\
$\mathrm{F_{cl}}$ &  0 &   6 &  0.063 &    0.10  &      1.7  &  $\infty$ \\
$\mathrm{G_{cl}}$ &  0 &   6 &  0.013 &    0.10  &      8.6  &  $\infty$ \\
$\mathrm{H_{cl}}$ &  0 &   6 &  0.003 &    0.10  &     43    &  $\infty$ \\
\hline\hline
\end{tabular}
\tablefoot{
The three subpanels refers to the 3-layer, 1-layer, and closed models
(from top to bottom respectively) in which boundary conditions are open (3-layer), half reflecting (1-layer
), and line-tied (close). 
For all runs:
$T_f=\infty$, $V_a^0=0.7~\mathrm{km/s}$,
$L_{ch}=2~\mathrm{Mm}$ (except for runs $H,~I,~L$ that have
$L_{ch}=1~\mathrm{Mm}$).
% Lengths in $Mm$, velocities in $km/s$.
$\epsilon$ denotes the ratio of photospheric over coronal Alfv\'en speed (it
plays no explicit role in the closed model). $L_c$ is the length of the coronal part of the loop. $l_\bot$ is the
perpendicular largest injection scale (the injected energy is distributed ar
scales $l_\bot$, $l_\bot/2$, and $l_\bot/4$). $U_0$ is the amplitude of the
input wave. $\chi_0$ is the linear to nonlinear time ratio (Eq.~\ref{chi_0}), $\chi_L$ is the leakage to nonlinear time ratio (Eq.~\ref{chi_L}).}
\label{table1}
\end{table}
%%%%%%%%%%%%%%%%%

Let us now give a brief account of the physical and numerical time scales.
Taking for instance $l_\bot=2~\mathrm{Mm}$, with $N=20$ \cha{perpendicular wave modes}, the largest available
perpendicular wavenumber will be $k_{max}=1.600~\mathrm{1/km}$.
The shortest nonlinear time (evaluated at the maximum perpendicular wavenumber in the corona) will be, if $z_c$ is the typical coronal amplitude (z denoting either $z^+$ or $z^-$)
\begin{equation}
\tau_{NL} = 1/(k_{max}z_c) \simeq \epsilon/(k_{max}U_0)
\end{equation}
where we have taken the resonant linear case (see next section) for which the wave amplitudes is larger by a factor $1/\epsilon$ in the corona.
Replacing by previous values and assuming $\epsilon = 0.01$ we obtain for the
smallest nonlinear time:
\begin{equation}
\tau_{NL} \simeq 6~10^{-5}~\mathrm{s}
\label{tnlnum}
\end{equation}
As a matter of comparison, we use $N_\parallel = 10^4$ grid points to describe
space along the loop, so that, for a typical loop length $L=6~\mathrm{Mm}$, we obtain for the smallest linear time for parallel propagation in the corona:
\begin{equation}
\tau_\parallel = 1/(k_\parallel^{max} V_A^c) = \epsilon L/(\pi N_\parallel
V_A^0) \simeq 2~10^{-3}~\mathrm{s}
\end{equation}
Hence, the constraint on the time step comes from the perpendicular nonlinear time.
Finally, at least in the linear case (see next section), the characteristic time for large scale evolution is the long leakage time
\begin{equation}
t_L = L / (V_a^0) \simeq 9~10^3\mathrm{s}
\label{tauA0}
\end{equation}
Comparing Eqs.~\ref{tnlnum}-\ref{tauA0}, we see that $\approx10^8$ time steps of a dynamical system with $ 2 \times 20 \times 10^4 $ degrees of freedom are necessary to achieve one (anticipated) characteristic evolution time of the system.
%To be complete, we must add the correlation time $T_f$ of the input spectrum,
%that will be most of the time infinite (or $3400$~s, see
%table~\ref{table1}). 
%%%%%%%%%%%%%%%%%%%%%%%%%%%%%%%%%%%%%%%
\section{Phenomenology}
%We consider here the zero-frequency case, in which the set ($z^+_1 , z^+_2 , ...z^+_N$) of amplitudes of the various modes input at the left foot point are time-independent.
%{\bf ca contredit pas mal de la discussion! On peut enlever ca? MV}
%%%%%%%%%%%%%%%%%%%%%%%%%%%%%%%%%%%%%%%
\subsection{Linear coronal trapping and leakage}
Let us first recall the linear result in the zero-frequency case, i.e.
when forcing is time independent: a transverse perturbation (here, any
perpendicular mode) is subjected to successive transmission-reflection at the two coronal
bases, left and right. Since nonlinear interactions are ignored, all modes show
the same evolution. As shown in \citet{Grappin_al_2008} for a loop with smooth variation of the Alfv\'en speed, the level of $z^+$ and $z^-$ grows progressively in the corona, in such a way as to achieve over a long time scale $t_L$ the asymptotic values
\begin{equation}
z^+ \simeq - z^- \simeq b/\sqrt\rho = U_0/\epsilon
\end{equation}
In other words, the asymptotic solutions are a 
uniform magnetic field amplitude everywhere along the loop 
at equipartition with the photospheric energy density, and, as well, 
a uniform velocity fluctuation everywhere along the loop. 
The asymptotic state is thus the same as
that would be achieved if the plasma were completely transparent to Alfv\'en waves ($\epsilon = 1$):
\begin{equation}
u=U_0, \ \	b=b_0 =B_0U_0/V_a^0 \label{asymp}
\end{equation}
although this happens on the long time scale $t_L= L /V_a^0$
and not on the short Alfv\'en coronal time  $t_a^c$.
(We assimilate here and in the following the coronal length to the total loop length L). 
Typically, if the shear amplitude is $U_0 = 0.1~\mathrm{m/s}$, and the mean
field $B_0 = 100~\mathrm{G}$, then the equilibrium magnetic field associated with the shear is the
equipartition field, that is, $b_0 \simeq 14.5$G.
%%%%%%%%%%
%
%One should remark that, within the same line, other, similar parameters may be defined. First, it may be more meaningful to use, instead of the photospheric nonlinear forcing time, to use the true coronal dissipation time $t_D$; we thus define the number $\chi_D$:
%\begin{equation}
%\chi_D = t_a^c/t_D
%\label{chi_D}
%\end{equation}
%However, the use of this number is limited because it is not known a priori: it is an output of the computation, not a control parameter.
%%%%%%%%%%
\subsection{Resonant response}
Consider the simplest case where the frequency of the photospheric input is either zero or resonant (that is, equal to $n/t_a^c$, with n an integer $\ge 0$). 
The coronal field perturbation induced by the photospheric field perturbation $U_0 = b_0/\sqrt \rho_0$ grows linearly with time until it saturates at a finite value because of the two damping losses, the linear leakage 
(with time scale $t_L$) and
the nonlinear turbulent damping (with time scale $t_D$):
\begin{eqnarray}
\partial_t b &=& B_0 U_0/L - b/t_L - b/t_D = B_0 U_0/L - b/t_\eta 
\label{dbdt0}\\
&=& b_0/t_L - b/t_\eta
\label{dbdt}
\end{eqnarray}
where 
$b_0=14.5G$ is the photospheric magnetic perturbation, and $t_\eta$ is the effective damping time:
\begin{equation}
t_\eta = (1/t_L+1/t_D)^{-1}
\end{equation}
Note that Eq.~\ref{dbdt} we have rewritten the first term 
using the definition $t_L=L/V_a^0$ in order to illustrate the fact that, in the absence of dissipation ($t_D=\infty,~t_\eta=t_L$) the trapping and leakage
times are equal.

The stationary solution is for the coronal field perturbation:
\begin{eqnarray}
b & = & U_0 \ (t_\eta B_0/ L)
\label{beq0}\\
  & = & b_0 /(1+t_L/t_D)
\label{beq}
\end{eqnarray}

One sees that the coronal response is maximal (equal to the photospheric value $b_0=14.5G$) when no turbulent damping is present ($t_D  \gg t_L$).
In the other limit ($t_L \gg t_D$), turbulent damping limits the coronal field
to a fraction $b_0$: $b \simeq b_0 t_D/t_L= t_D B_0 U_0/L$.

Relation~\ref{beq} may be rephrased in terms of energy per unit mass as
\begin{eqnarray}
E = E_0 (t_\eta/t_a^c)^2
\label{res}
\end{eqnarray}
with $E_0=1/2(U_0^2+b_0^2/4\pi\rho_0)=U_0^2$.
In the case where $t_D \ll t_L$, Eqs.~\ref{beq}-\ref{res} have been already given by \citet{Hollweg_1984a}; as
remarked by \citet{Nigro_al_2008}, they are valid also for the zero frequency case (see
also \citealt{Grappin_al_2008}), the only difference being that in the latter case magnetic energy is dominant in the corona, while in the case of non zero resonance 
coronal magnetic and kinetic energies are at equipartition.

A last remark concerns the use of Eqs.~\ref{dbdt}~and~\ref{beq} (but
not Eq.~\ref{res}). 
Caution must be taken when applying the line-tied limit, 
$t_L=\infty$, since the trapping time, appearing as $t_L$ in
these equations, is finite and fixed. Hence the explicit forms,
Eqs.~\ref{dbdt0}~and~\ref{beq0}, are better suited to understand the difference between the opened and closed models. 
In particular one sees that the coronal magnetic field grows linearly with time
in absence of dissipation (Eq.~\ref{dbdt0}) while, when dissipation is present,
it can grow well beyond the
leakage-limited value $b_0$ (Eq.~\ref{beq0}), since the loss timescale $t_\eta$
has no upper limit\footnote{As we will see in the closed
model the dissipation timescale can be larger than the nominal value $t_L$ 
(see Fig.~\ref{triple-ter}), thus leading to $b>b_0$.}

\subsection{The general case}
In general, the signal injected into the corona is not necessarily resonant
\cha{and more generally not monochromatic}. To quantify both the trapped energy and its dissipation rate we need to know how the \cha{time-dependent} energy input is distributed between resonant and non-resonant frequencies.
We thus introduce the correlation time
of the energy $T_{cor}^0$ entering the corona or equivalently the width of the
injection spectrum $1/T_{cor}^0$, which is \cha{a priori} unknown.
\begin{figure}
\begin{center}
\includegraphics [width=\linewidth]{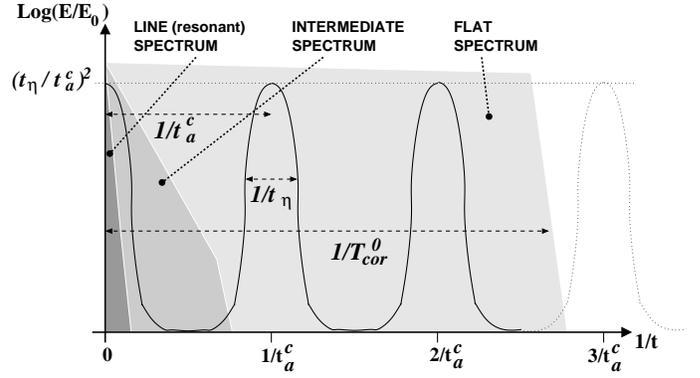}
\caption{
Sketch of the linear coronal energy gain, $\log(E/E_0)$ as a
function of frequency. $E_0$ is the input photospheric energy at each frequency. It is assumed that $t_a^c<<t_\eta$, and only the non-wkb portion of the spectrum is shown.
The injected spectrum is also indicated as a shaded areas in arbitrary scale, for the case of flat, intermediate, and line spectrum
(in increasing gray scale order). If leakage is neglected
($t_L\to\infty$) dissipation determines the height and width of resonances,
since it is the only mechanism that limits the energy accumulation. For weak
dissipation (and generally for $t_\eta\to\infty$) the resonances tend to delta functions.
}
\label{fig:spect}
\end{center}
\end{figure}

Recalling that the resonant lines are spaced each $1/t_a^c$, each with a width
equal to the inverse of the damping time $t_\eta$, then we may distinguish
several cases depending on the portion of the excited spectrum (see Fig.~\ref{fig:spect}):
\begin{itemize}
\item Flat spectrum : $T_{cor}^0 < t_a^c$.
Since negligible energy is transmitted outside the resonant lines (enlarged by
damping) compared to the energy transmitted for frequencies within the lines
(anti-resonances),
this leads to a filling factor equal to $t_a^c/t_\eta$ compared to a spectrum
\cha{made} of only resonant \cha{frequencies} (Eq.~\ref{res}).
\item Intermediate $T_{cor}^0$ : $t_a^c < T_{cor}^0 < t_\eta $.
Then the filling factor is $T_{cor}^0/t_\eta$ as only the zero\cha{-frequency} resonance and
the first anti-resonance are excited.
\item Long correlation time or resonant spectrum: $t_\eta < T_{cor}^0$.
\cha{This} coincides with the linear resonant gain
Eq.~\ref{asymp} if $t_L<<t_D$.
\end{itemize}
Finally:
\begin{eqnarray}
E = E_0 (t_\eta/t_a^c) &\;&\;(T_{cor}^0 < t_a^c)\;\; 
\label{chi1} \\
E = E_0 (t_\eta/t_a^c) \ T_{cor}^0/t_a^c  &\;&\; (t_a^c < T_{cor}^0 < t_\eta)  \;\; \label{chi2}\\
E = E_0 (t_\eta/t_a^c)^2 &\;&\; (t_\eta < T_{cor}^0) \;\;
\label{chi3}
\end{eqnarray}
Note that we transformed Eq.~\ref{chi2} which originally reads 
$E = E_0 (t_\eta/t_a^c)^2 \ T_{cor}^0/t_\eta$.
Equations~\ref{chi1}-\ref{chi2} have been derived for negligible leakage
($t_\eta=t_D$), in the strong turbulence case by \citet{Hollweg_1984a} and in
the weak turbulent case by \citet{Nigro_al_2008}. The relations proposed here
\cha{extend these early findings by including} the case where leakage dominates turbulence, and the case of very weak
turbulence (resonant spectrum).

To make these expressions explicit, one should express the unknown parameters in terms of control parameters.
It is tempting for instance to identify $T_{cor}^0$ with $t_{NL}^0$: then
the three regimes correspond respectively to strong turbulence ($\chi_0>1$),
weak turbulence ($\chi_0<1$), and weak dissipation ($\chi_L<1$).
\begin{figure*}[t]
\begin{center}
\includegraphics [width=\linewidth]{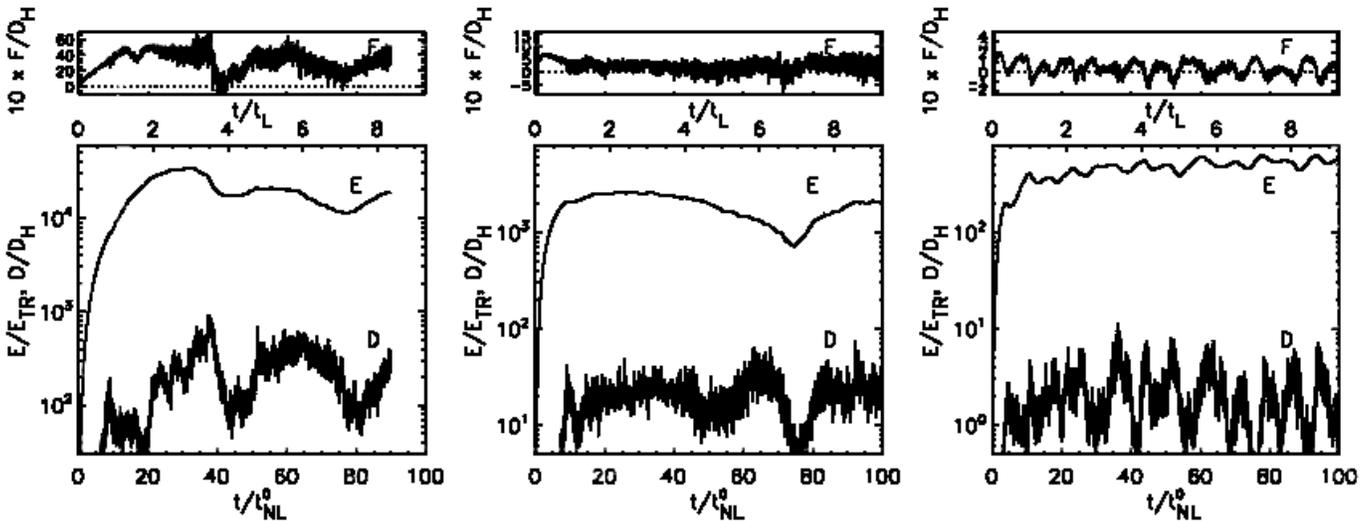}
\caption{From left to right:
Run $\mathrm{D_{cl}}$ (closed model), Run $\mathrm{D_{1L}}$ (1-layer model) and
Run D (3-layer model).
For all the runs $\chi_0\approx0.2$, for the
1-layer and 3-layer model $\chi_L\approx 11$.. 
\textsl{Top panels}: time evolution of the net energy flux
$F$. \textsl{Bottom panels}: time evolution of the coronal energy ($E$) and
dissipation below ($D$). Time is normalized to the input nonlinear 
timescale $t_{NL}^0$, energy is normalized to the injection energy at the left
boundary $E_{TR}=U_0^2$, dissipation and flux are normalized to Hollweg
expression $D_H=E_{TR}/t_a^c$.}
\label{fig3}
\end{center}
\end{figure*}

Indeed, the latter condition can only be satisfied if $t_L<t_D$ 
\cha{if we exclude the possibility that 
$t_D<t_{NL}^0$}. 
Thus assuming
$T_{cor}^0=t_{NL}^0$, the line spectrum coincides with the  
linear resonant gain, and can only be reached \cha{by} imposing $\chi_L<1$. 
If the injection spectrum has a finite width, a possible 
choice \cha{is} $T_{cor}^0=\mathrm{min}(T_f,t_{NL}^0)$ as suggested by
\citealt{Malara_al_2010}. If $T_f>t_{NL}^0$ we fall in the previous case. 
If instead $T_{f}<t_{NL}^0$, the ordering considered in \citet{Nigro_al_2008,
Malara_al_2010}, the line spectrum is not achievable. However, as we will see,
the correlation time may also be given by other timescales, \cha{like} the leakage
time $t_L$ and/or the chromospheric crossing time
$t_a^{ch}=L_{ch}/V_a^0$.
The remaining (difficult) task is to express the dissipation time $t_D$ (since
$t_\eta=min(t_D,t_L)$), in terms of $\chi_0$ and $\chi_L$ via the coronal nonlinear time. 
We will come back \cha{to} this point later on.

\subsection{Dissipation in the strong and weak regimes}
In the strong \cha{turbulence} case ($\chi_0>1$) dissipation is expected to dominate
on leakage and a simple explicit expression of the
dissipation rate is obtained after replacing $t_\eta=t_D$ in 
Eq.~\ref{chi1} \citep{Hollweg_1984a}:
\begin{equation}
D = E/t_D = E_0/t_a^c
\label{hol}
\end{equation}
This relation is attractive, as it leads to a universal result: the heating rate per unit mass does not depend on the detail of turbulent dissipation, since it depends only on the length of the loop and the photospheric energy.
However, this universality is lost when we turn to the weak turbulent regime, $\chi_0 < 1$, Eq.~\ref{chi2},
which we have seen is probably prevalent in the corona (Fig.~\ref{fig1a}).
To extrapolate the previous expression (Eq.~\ref{hol}) to the weak regime with $\chi_0 < 1$, we identify $T_{cor}^0$ with $t_{NL}^0$ in Eq.~\ref{chi2} and still adopt $t_D < t_L$:
\begin{equation}
D = E/t_D = E_0/t_a^c \ (1/\chi_0)
\label{holweak}
\end{equation}
This predicts that, the weaker the turbulence regime, the higher the dissipation.
We will see that both relations~\ref{hol}-\ref{holweak} are reasonably satisfied if we use the line-tied limit, but not in the more realistic open case. 
In the open case, we will find that actually Hollweg's expression
(Eq.~\ref{hol}) holds more or less both for $\chi_0 >1$ and $\chi_0<1$, which requires to admit that $t_D > t_L$ in the weak regime, i.e., that the dissipation time becomes very long as turbulence becomes weaker.

%-----------
\section{Results}
In the following we will compare first the different models in a weak
turbulence case, the most probable for coronal conditions. 
Then we will focus on the 3-layer model, comparing the weak and strong turbulence regimes.
\subsection{How leakage changes turbulence: the weak turbulence case}
We consider here a weak turbulent case with $\chi_0\approx0.2$, and 
compare the closed, 1-layer, and 3-layer models.
The runs are $\mathrm{D_{cl},~D_{1L},~and~D}$ respectively in Table~\ref{table1};
in the open models 
$\chi_L\approx 11$ so we expect that turbulence is the main channel
for energy loss in all models.
Because of this, we should not expect significant differences between the closed 
and the 1-layer run. However, we might perhaps find differences due to the different 
forcing (from now on we will use forcing to mean injection into the corona)
between the 1-layer and the 3-layer runs, recalling that forcing is constant in the first
case, and time-dependent in the second, due to the possibility of a chromospheric turbulence.
 
The time evolution of the corona in the three models (from left to right) is summarized in Fig.~\ref{fig3} 
where the entering energy flux $F$ (top panel), the total energy $E$ and dissipation $D$ (bottom panel) are shown
(see Eqs.~\ref{fluxa}-\ref{fluxc}).
Time is normalized to the input nonlinear timescale, $t_{NL}^0$, energy is
normalized to the input coronal energy $E_{TR}\equiv z_{TR}^2/4$, the
dissipation and the flux are normalized with respect to Hollweg expression, 
$D_H\equiv E_{TR}/t_a^c$ (for the closed and 1-layer model $z_{TR}\equiv U_0$,
for the 3-layer model $z_{TR}$ is the measured quantity $z^+_1$ that is not directly controlled by the boundary conditions).
%-----------
\begin{figure}[th]
\begin{center}
\includegraphics [width=\linewidth]{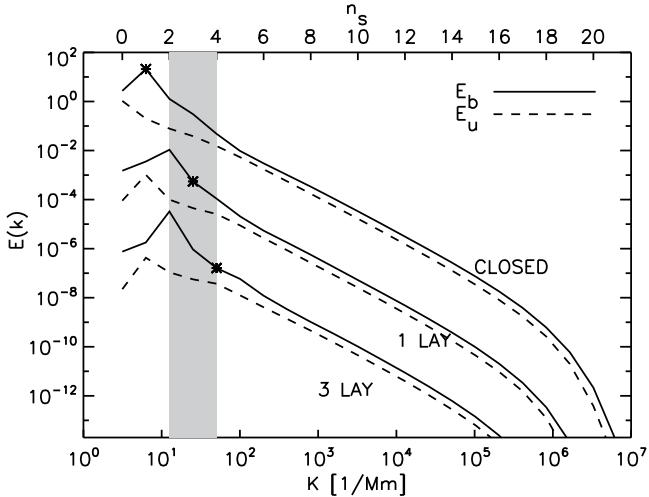}
\caption{Coronal kinetic energy spectrum (dashed line) and magnetic energy spectrum (solid line) for runs $\mathrm{D_{cl},~D_{1L},~and~D}$.
Wavenumbers are in units of 1/Mm, in the top x-axis the corresponding shell
numbers are indicated. The spectra are averaged in time and space, the
normalization is in arbitrary units (spectra are also rescaled) . The symbols on the $E_b$ spectra indicate the first shell number for which $t_{NL}^u(k)=1/ku(k)<t_a^c$.}
\label{fig4-1}
\end{center}
\end{figure}

A quick look at the energy flux curves shows a sharp contrast between
the closed run and the open 1-layer run.
While in the closed case, the coronal energy flux is almost always positive, in the open case, it is constantly oscillating around zero, although with a positive mean value flux.
This has an immediate corollary: the energy level shows much lower values in the open case.
Another corollary is that the dissipation rate itself, i.e., coronal heating, is reduced by a factor ten.
This tendency is sharply enhanced in the case of the three-layer model
which shows a further reduction of a factor 5.
Another remarkable difference appears in the 3-layer model, which
accounts for the chromospheric turbulence. The energy and the energy
flux display quasi-periodic oscillations that are absent in the closed and
1-layer model whose energy time-series are shaped
by the time-independent forcing.
Such oscillations have a periodicity of the order of one leakage time or smaller
(see the top horizontal axis in the bottom panel). However we cannot rule out
that their origin lies in the chromospheric turbulence. Indeed, the
periodicity happens to be close to two chromospheric crossing times
$2t_a^{ch}=2L_{ch}/L_c t_L=2/3 t_L$, which we interpret as the timescale
necessary for waves injected from the left footpoint to leave the chromospheric layer (a round trip of the chromosphere). Most probably such
oscillations are due to the coupling of the chromospheric and coronal
turbulence and both timescales matters, as we will see in 
section~\ref{weakstrong}.

\begin{figure}[t]
\begin{center}
\includegraphics [width=\linewidth]{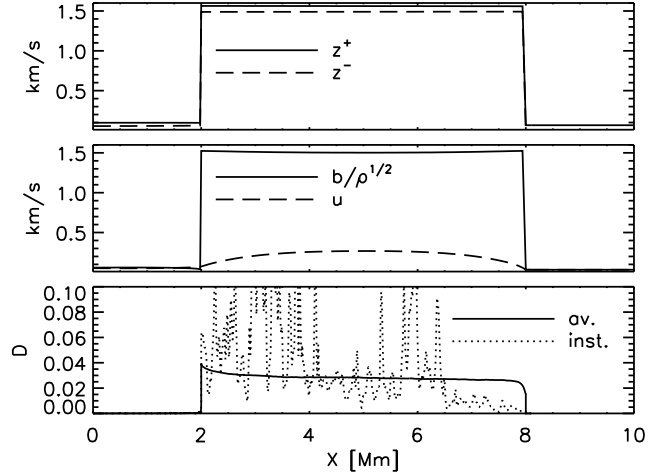}
\caption{Run D (3-layer open model, weak turbulence case): spatial
distributions of fluctuations (top and mid panels) and  
turbulent heating (bottom panel). The time averaged rms amplitude (in
km/s) are plotted as a function of loop coordinate (in Mm) for $z^+,~z^-$ (top panel, solid and dashed line respectively) and for $b/\sqrt{4\pi\rho_c}, u$ (bottom
panel, solid and dashed line respectively). The time-averaged heating rate (in
arbitrary units) is
plotted in the bottom panel as a solid line, also a snapshot is shown in dotted line.
}
\label{fig3lay}
\end{center}
\end{figure}
%-----------
Fig.~\ref{fig4-1} shows the time and space-averaged kinetic and magnetic spectra in the corona for the three models. One sees that all cases show well-developed power-law ranges, 
plus a magnetic hump at large scales. 
The (common) forcing range is represented by a gray vertical band.
Note also the symbol on the magnetic spectrum which marks the large scale for which the
effective nonlinear time computed on the rms velocity at that scale is smaller than the Alfv\'en crossing time.
The only significant difference visible between the three spectra is that the magnetic peak is located at the largest forcing scale for the open runs, while it has migrated to a scale larger by a factor two for the closed
run.
This indicates that an inverse transfer is active in all cases, but that it is more active in the closed
case, or also possibly that it has been hindered by leakage of the largest scales in the open cases.

We are thus forced to conclude that, in the open models, despite the fact that
$\chi_L>1$, the energy accumulation is limited by leakage.
This means that the nonlinear timescale $t_{NL}^0$ is a sharp under-estimation of 
the real dissipation timescale. We will come back on this point in the following.

%-----------
\subsection{The 3-layer model: chromosphere vs corona}
We detail here the structure of the open 3-layer model in the weak turbulence case. 
In particular we compare the chromosphere and corona.
We show in Fig.~\ref{fig3lay} the spatial profiles in the corona and
chromosphere of the fluctuations and of the dissipation rate.
The top panel shows the time average of the rms value $z^+$ and $z^-$
amplitudes with $z^\pm_{rms}$ defined as:
\begin{equation}
z^\pm_{rms}=\sqrt{\sum_n |z^\pm_n|^2}
\end{equation}
The mid panel shows the time averaged rms values of velocity $u$ and magnetic
field in km/s units ($b/\sqrt{\rho})$.
The bottom panel shows the time average (solid line) and a snapshot (dotted
line) of the heating rate.

In spite of the presence of turbulence (as revealed by the spectra examined above), 
the rms amplitudes of all quantities are seen to be remarkably smooth functions of 
loop coordinates except of course at the T.R.. 
Main features are :
(1) the magnetic field amplitude in the corona and chromosphere are actually comparable
(the magnetic field amplitude plotted in the figure is $b/\sqrt{4\pi\rho}$, hence a factor of about $1/\epsilon= 50$ between the coronal and chromospheric values);
(2) the velocity contrast is significantly larger than unity but much smaller
that the magnetic contrast (in units of velocity), and its coronal profile has a simple form
(3) the $z^+$ and $z^-$ levels are comparable in the corona
%--------------
\begin{figure}[t]
\begin{center}
\includegraphics [width=\linewidth]{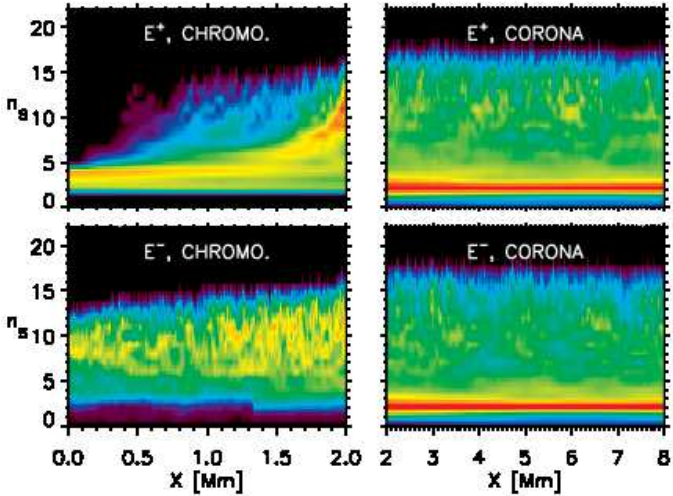}
\caption{Run D. Contour plot of the spectra $E^\pm(x,k_n)$
(snapshots) for $z^+$ and $z^-$ (top and bottom panels respectively)
compensated by $k^{5/3}$ in the chromosphere (left
panels) and in the corona (right panels). Ordinate: shell number
$n_s=\mathrm{log}_2(k_n/k_0)$. Abscissa: coordinate $x$ along the loop in Mm. The contours have different ranges in the chromospheric and coronal
layers to better highlight their structures.
}
\label{figEkxchr}
\end{center}
\end{figure}

(1) implies that the main part of the magnetic energy trapped in the corona is actually close to the linear state of zero resonance  (in the linear case with zero frequency the asymptotic coronal magnetic field fluctuations is equal to the photospheric field, see Grappin et al 2008).
(2) the coronal profile of the velocity field is actually close to the profile of the first linear resonance   \citep{Nigro_al_2008}.
(3) allows full nonlinear coupling which is compatible with the existence of a developed spectrum.

\cite{Nigro_al_2008} had already found in the closed case that the characteristic linear resonance profiles of the coronal cavity are not deeply affected by the presence of a nonlinear cascade.
It appears that the same linear resonance profiles are'nt affected by leakage either.

Finally, the time-averaged profile of the average dissipation rate per unit mass (bottom panel in Fig.~\ref{fig3lay}) shows that the chromospheric dissipation remains negligible,
and also that the left T.R. (i.e. above the foot point where energy is injected)
is dissipating at a slightly higher rate than the other foot point.
A typical snapshot (in dotted line) is also shown, providing a hint of the
substantial intermittency of the heating rate, both in space and time.
%--------------
\begin{figure*}[t]
\begin{center}
\includegraphics [width=\linewidth]{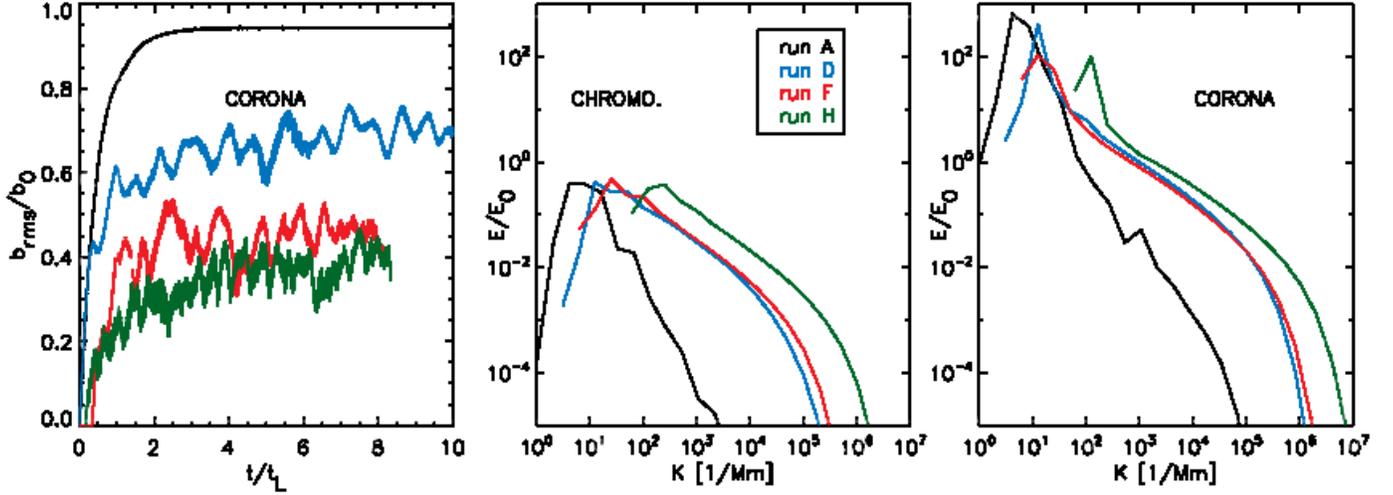}
\caption{
Runs $A$, $D$, $F$, and $H$ with increasing $\chi_0$.
%energy growth and spectra.
Left panel: growth of rms coronal magnetic field (normalized to its asymptotic
linear value $b_0=15$~G). Middle and right panels:
total energy spectrum $E(k)$ averaged in time and space in the chromosphere
and in the corona respectively.}
\label{fig4a}
\end{center}
\end{figure*}
%--------------------------

The turbulent activity of both the chromosphere and corona are shown in figure~\ref{figEkxchr}, in which we plot snapshots of the $z^-$ and $z^+$
spectra $E^\pm(x,k_\bot)$ (top and bottom panels) respectively in the left chromosphere and corona
(left and right panels).
The spectra are compensated by $k^{5/3}$ in both layers.
Note that the motivation for plotting $z^\pm$ spectra instead of $u$ and $b$ spectra is
to make clear the respective contributions of the chromosphere and corona to the 
spectral formation, as the directions of propagation are identifiable for $z^\pm$, not for $u$ and $b$. Note that only the left chromosphere has been represented (since in the right chromosphere the evolution is purely linear, due to the absence of $z^-$ input from the right foot point), 
that its length has been enlarged to make its structure more conspicuous, and
that the contours have different ranges in the chromopshere and in the corona.

One can see in the figure something like the trajectory of turbulence from the left foot point to the corona and all the way back (so, one begins from the top left panel and proceeds clockwise).
First, in the chromosphere the onset of turbulence does not take place immediately
starting from the left foot point : the $z^+$ spectrum (top left) first shows
only the 3 injected scales (seen as a red-yellow band), and only very progressively adds smaller
scales (first seen as a blue haze).
Spatial intermittency then appears about in the middle of the chromosphere in the form of
small scale filamentary structures.
Note that this corresponds to a travel time $\Delta L/V_a^0 \simeq 1100$~s, which is close to the nonlinear time $t_{NL}^0 \simeq 800$~s.

In the corona (top panels) one sees on the contrary no large parallel gradients:
\cha{as was seen previously with the rms $z^+$ and $z^-$ energies.}
\cha{A} conspicuous feature of the coronal spectrum is the hump appearing as a red ribbon
which is displaced towards large scales (when compared to the peak in the chromospheric
injected spectrum, top left).
This again reveals the inverse transfer already noted above in Fig.~\ref{fig4-1}.

Finally, one sees in the bottom left panel that the wave leaking from the corona makes
the $z^-$ chromospheric spectrum look much more developed than its $z^+$ counterpart.

\subsection{The 3-layer model: Increasing turbulence}\label{weakstrong}
We now increase in the 3-layer model the turbulence strength $\chi_0$ from $0.04$ to $4.8$ 
(runs $A$, $D$, $F$, $H$).
This is achieved by decreasing the nonlinear time, while the leakage time is fixed.
Hence even though we have already seen that the nonlinear time is clearly a strong
lower bound for the dissipative time, again, one should expect 
the open model to match at some point the closed model in the limit $\chi_L \gg 1$. 
This point will be considered again in the discussion where the properties of all models are summarized.

In Fig.~\ref{fig4a} we illustrate how the dynamics changes in the open 3-layer model when increasing $\chi_0$. The left panel shows the rms magnetic field amplitude in the corona normalized to $b_0$, the linear zero frequency solution, while the two other panels show the (space and time averaged) total energy spectra respectively in the chromosphere and the corona.

The main points are
(1) When the nonlinear time is too large (very small $\chi_0$, run A), one sees that 
turbulence has no time to develop before reaching the corona.
Both the chromospheric and coronal spectra remain largely devoid of small scales. 
Dissipation is thus negligible.
The asymptotic level of the magnetic field is close to its $15$ G linear value, the growth of $b_{rms}$ being extremely regular and devoid of any small scale fluctuations. All this happens
in a leakage time.
(2) Decreasing the nonlinear time progressively decreases the asymptotic coronal field.
Its growth becomes now chaotic, the signal in the left panel showing a whole spectrum of frequencies, with
most remarkably periods close to the leakage time for the two intermediate
values of $\chi_0$,
but also periods close to two chromospheric crossing time
$2t_a^{ch}=1/3 t_L$ for the strongest $\chi_0$ (run $H$ in
the left panel, see for example the range $t/t_L\in[3,4]$).
(3) At reasonably large $\chi_0$, the coronal spectra are developed. However, the chromospheric spectra are significantly steeper.
In the chromosphere, the slope is close to $1.8$, while it is close to $1.7$ in the corona.
(4) Note also that the chromospheric spectra are devoid of the humps which appear in the coronal spectra.
\begin{figure}[t]
\begin{center}
\includegraphics [width=\linewidth]{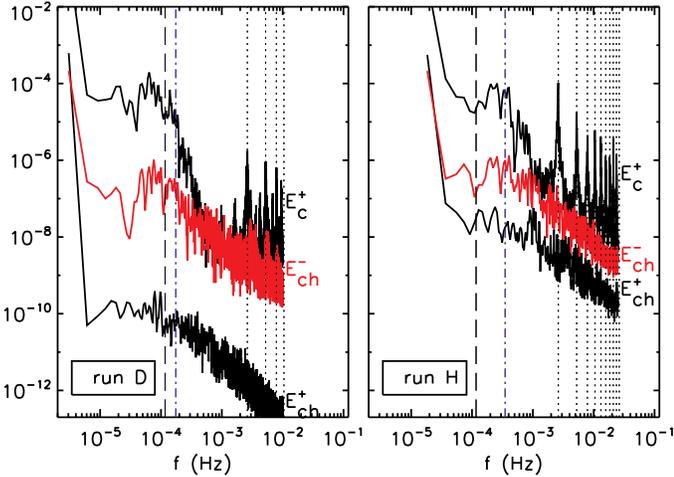}
\caption{
Runs $D$ and $H$ (weak and strong $\chi_0$): Frequency energy spectra $E^+(f),~E^-(f)$,
computed by taking the Fourier transform of the $(z^\pm_{rms}(t))^2$ at each plane and then space averaging separately
in the chromosphere (bottom black and red lines) and corona (top black line).
The $E^-$ (not plotted) and $E^+$ coronal spectra are indistinguishable. The vertical lines mark some relevant timescales: dotted lines for the resonances at $n\ge1$, dotted-dashed lines for the (round-trip) chromospheric crossing time, $2t_a^{ch}$, and long-dashed lines for the leakage time, $t_L$.}
\label{fig5a}
\end{center}
\end{figure}

We thus conclude that too weak a cascade does not change linear zero frequency results at all, and
that there is a $\chi_0$ threshold above which turbulence has common properties. 
There are slight differences 
in the chromosphere and corona, the main ones being the large scale coronal peak, and a slightly different slope.

We now examine frequency spectra. 
We have computed frequency spectra
 of $(z^\pm_{rms})^2$ at each position along the loop; 
 then they have been separately averaged in the corona and in the chromosphere. 
The original time series has been windowed with the hanning procedure
and the zero frequency is also displayed as the lowest frequency in the plot (Fig.~\ref{fig5a}). 
The coronal $z^+$ and $z^-$ have practically the same spectrum, so we plot 
only $z^+$ in the corona, as well as the chromospheric spectra of $z^+$ and $z^-$. 

One sees on the coronal spectra the appearance of peaks close to (but not coinciding exactly with) the resonant frequencies: $n /t_a^c$, these harmonics being marked as dotted lines.
This confirms again that the quasi-linear trapping properties are not strongly affected by leakage.

For weak turbulence (left panel)
the spectra are dominated by the lowest frequencies, the input zero-frequency and a low-frequency bump. As the strength of turbulence is increased (right panel), more energy goes into finite-frequency resonances, some of them becoming as
energetic as the low frequency part of the spectrum. Note that the location
of the low-frequency bumb corresponds roughly to two characteristic timescales,
the leakage time, $t_L$, and two chromospheric crossing time, $2t_a^{ch}$,
which we interpret as the signature of the turbulence activity in the coronal and
chromospheric layer respectively. 
In run $D$ (low turbulence) two distinct bumps appear in the chromopheric
spectrum $E_{ch}^+$ at frequencies $1/t_L$
and $1/2t_a^{ch}$, while in run $H$ (strong turbulence) the bump lies in between
them. In the coronal spectra (and also in $E_{ch}^-$) the bump is somewhat
wider, possibly showing a coupling with the (zero and finite frequency) resonances.
The importance of both timescales points out the fact that the low-frequency spectrum in the 3-layer
model is affected by the coupling between the coronal and chromospheric turbulence.
%-------------------------------
\section{Discussion}
\begin{figure}[t]
\begin{center}
\includegraphics [width=\linewidth]{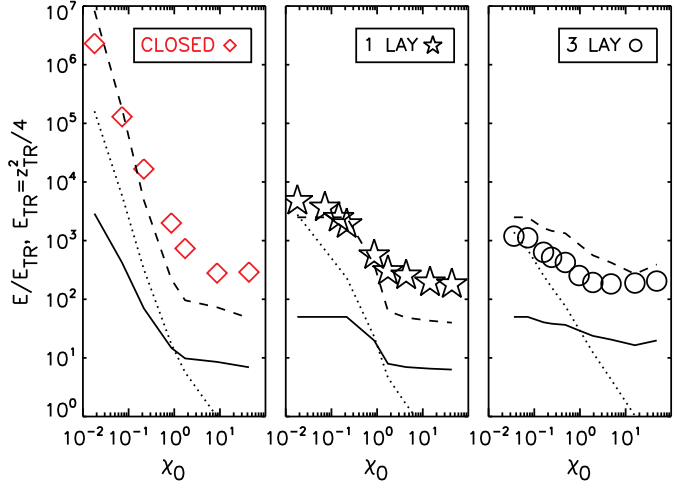}
\caption{
Coronal energy per unit mass normalized by the T.R. injection energy as a
function of the time ratio $\chi_0$. 
Left: closed model results (symbols);
Mid panel: 1-layer model results (symbols);
Right: 3-layer model results (symbols);
dashed line: resonant scaling; dotted line: intermediate correlation time scaling; 
solid line: small correlation scaling.}
%The vertical dotted lines mark the values $\chi_0=\epsilon$ and
%$\chi_0=1$.}
\label{fig-ene}
\end{center}
\end{figure}

We have studied in this work the problem of heating coronal loops by
forcing a photospheric shear through the injection of
Alfv\'en waves at one foot point, and waiting for the injected Alfv\'en waves 
to be transmitted into the corona, to be trapped (amplified), and finally to
dissipate due to turbulence.
We used for this a simplification of the RMHD equations that exploits Shell models to account for the perpendicular nonlinear coupling.

In contrast to previous work, we have considered a finite Alfv«en speed
difference between the photosphere and the corona and freely propagating waves
deep in the photosphere, thus allowing energy to leak back to the chromosphere
and to deeper layers of the solar atmosphere.
We found that, although leakage doesn't change
dramatically either the quasi-linear trapping properties of the corona or the spectral properties of turbulence, it alters strongly the level of the energy trapped in the corona and the resulting dissipation. 

We have seen that the coronal energy and dissipation rate vary, depending (i)
on the ratio $\chi_0$ of the linear Alfv\'en crossing time 
by the nonlinear input time $t_{NL}^0$ (ii) on the choice of the atmosphere
model, allowing or not the development of a chromospheric turbulence.
We now present systematically these variations and discuss them.
We will present on an equal footing the closed model, 1-layer and 3-layer
models.
The closed model results should allow direct comparison with earlier
work, and the comparison between the 1-layer and 3-layer models should make clear the effect of chromospheric turbulence.
However, we consider that the 3-layer model is the more realistic of the three
models.

Note also that we will vary $\chi_0$ while fixing the Alfv\'en speed contrast
to $\epsilon=0.02$ and the loop legnth to $L_c=6~\mathrm{Mm}$.
The dependence on $\epsilon$ and $L_c$ will be postponed to a further study.

%\uwave{We have seen that the coronal energy and dissipation rate vary, depending (i) on the ratio $\chi_0$ of the linear Alfv\'en crossing time by the nonlinear input time $t_{NL}^0$ (ii) on the choice of the atmosphere model, allowing or not the development of a chromospheric turbulence.
%We now present systematically these variations and discuss them.
%We will present on an equal footing the closed model, 1-layer and 3-layer models, in the hope to make the origin of the variations clearer, although in our opinion the 3-layer model is by far 
%the more realistic of the three models.
%In particular, the closed model results should allow direct comparison with earlier work.
%Note also that we will vary $\chi_0$ while fixing the Alfv\'en speed contrast to $\epsilon=0.02$. 
%The dependence on $\epsilon$ will be postponed to a further study.
%}

\subsection{Energy}
We show in Fig.~\ref{fig-ene} how the energy trapped in the corona depends on the turbulence strength $\chi_0$ in the different models (left: closed model, mid panel: 1-layer, right: 3-layer).
In each case, we have shown how the results are fitted by the generalization of
the Hollweg-NMV model as given by the three different possible regimes
(Eqs.~\ref{chi1}-\ref{chi3}). 
For this purpose we normalize the energy to the
coronal input energy $z_{TR}^2/4$ which coincides with the input energy $U_0^2$
in the closed and 1-layer model while it is a measured quantity for the 3-layer
model ($z_{TR}^2\equiv|z^+_1|^2$ in Fig.~\ref{fig1}).

The regime is identified by the choice of the correlation time of the input spectrum, which is different in the different models. 
For instance, a constant signal in the closed model is a particular case of the resonant regime (dashed line, line spectrum). The flat spectrum should be found in the case of a very short correlation time (solid line), and the intermediate regime by the dotted line.
It is striking that the best fit, although largely imperfect, is always obtained by the resonant
expression (zero-frequency spectrum), whatever the turbulence strength (i.e.
both for small and large $\chi_0$).

The behavior of the system is therefore dominated by the $n=0$
resonance. This could be somewhat expected for the closed and 1-layer models,
since the input signal is time independent, but less so for the
3-layer model, in which the chromospheric turbulence modifies the
input frequency spectrum to the corona, at least for large $\chi_0$.
Hence, it seems that the dominance of the low-frequencies is not
caused by the particular forcing chosen here, but it is a consequence of the
coronal activity itself. 
As we have seen, the coronal spectrum shows a bump at large perpendicular
scale, containing very low frequencies; this bump is not present in the
chromospheric spectrum (see fig~\ref{fig4a}).

A last point concerns the model differences. 
While all three models attain the same energy level for $\chi_0>1$, it is seen that they strongly differ in the weak turbulence regime.
In the closed model the energy grows until turbulence becomes efficient enough to balance the input energy. In the opened models, instead, the energy accumulation is prevented by the
leakage and the resulting level is much lower.
\begin{figure}[t]
\begin{center}
\includegraphics [width=\linewidth]{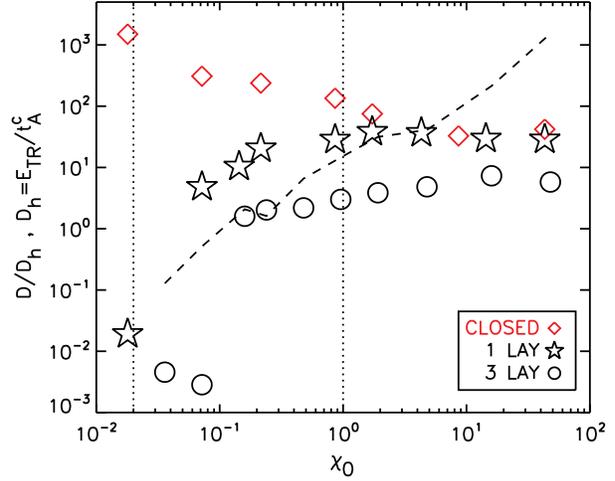}
\caption{
Dissipation normalized by Eq.~\ref{hol} vs $\chi_0$
for the three models.
The vertical dotted lines mark the values $\chi_0=\epsilon$ ($t_{NL}^0=t_L$) and $\chi_0=1$.
The dashed line is the prediction from Eq.~\ref{SUPER} for the 3-layer
model.}
\label{triple-bis}
\end{center}
\end{figure}

\subsection{Heating rate}
The heating rate is shown in a single plot for all three models in Fig.~\ref{triple-bis}.
We show the dissipation per
unit mass normalized to Hollweg expression, $D_h=E_{TR}/t_a^c$,
(Eq.~\ref{hol}) vs $\chi_0$:
diamonds are for the closed model, stars for the 1-layer model, and circles for the 3-layer model.

In the weak turbulent part of the diagram ($\chi_0 < 1$), we find again the same ordering of the models observed above for the energy:
a line-tied model leads to a heating rate inversely proportional to the turbulence strength
(our results follow well the fit $D \propto (u^2/t_a^c) \chi_0^{-1/2}$ proposed
by \citealt{Dmitruk_99})
while for the open models the heating rate goes down proportionally to the turbulence strength.
We note however the sudden drop of the dissipation rate at very low
$\chi_0$ (roughly corresponding to $t_{NL}^0>t_L$, their equality
being indicated by the dotted vertical
line at $\chi_0=\epsilon$), due to the
absence of formation of a high-wave number spectrum in the open models:  
the coronal energy level is too low to trigger a cascade before fluctuations leak out of the corona.
 
In the strong turbulent regime ($\chi_0>1$) the dissipation is about independent
of the turbulence strength: the level of this plateau is common to the closed
and 1-layer models which differ only in their boundary conditions, but it is
lower for the 3-layer model. 
This is to be attributed to the chromospheric
turbulence that reduces the coronal input and hence the actual strength of the
turbulence, for given $\chi_0$.
\begin{figure}[t]
\begin{center}
\includegraphics [width=\linewidth]{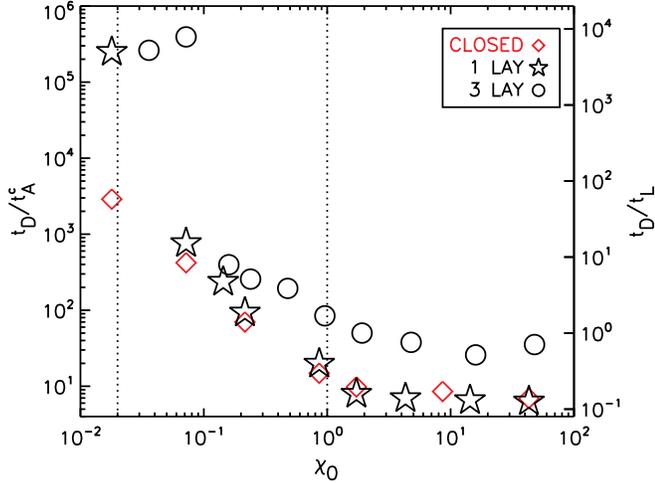}
\caption{
Dissipation time vs $\chi_0$ for the three models;
note the different normalization in the left and right axis.
The vertical dotted lines mark the values $\chi_0=\epsilon$ 
($t_{NL}^0=t_L$) and $\chi_0=1$.}
\label{triple-ter}
\end{center}
\end{figure}

\subsection{Dissipation time}
Finally we consider in Fig.~\ref{triple-ter} the dissipation time,
$t_D=E/D$, normalized to the coronal crossing time
$t_a^c$. 
For the opened models the right vertical axis also shows the dissipation time normalized to the
leakage time, which makes sense since the ratio $\epsilon$ between the Alfv\'en and leakage time
remains fixed in the data shown here.

The few points of the open models with very low turbulence strength $\chi_0$ show very high values of the dissipation time, due to the undeveloped turbulence, as already mentioned above.

Leaving apart these points, one sees that increasing $\chi_0$ (i.e., decreasing the input nonlinear time), the dissipation time decreases as expected, but that, most remarkably, it stops decreasing when $\chi_0$ reaches about unity.
The plateau is comparable for the 1-layer and closed models, corresponding to $t_L \simeq 10 t_D$ in the 1-layer model, while it corresponds to $t_L \simeq t_D$ in the 3-layer model.

It would certainly be a progress, both from the theoretical and the practical viewpoint,
to understand how the heating time and energy are related.
The difficulty is that the usual relations here are modified by the existence of the large-scale hump in the spectrum: one cannot consider that there is a straightforward (direct) cascade from the large to the small scales, as the energy is clearly "blocked" at large scales.
\begin{figure}[t]
\begin{center}
\includegraphics [width=\linewidth]{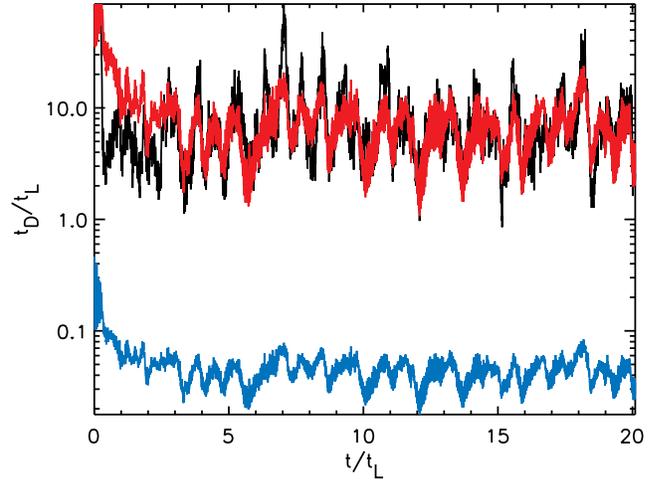}
\caption
{Run D: Comparing the instantaneous dissipation time with the slow Iroshnikov-Kraichnan time scale (red) and the standard nonlinear time (blue, below).
Note that the agreement between the measured dissipation time and IK time is as good as here for all values of the turbulence parameter $\chi_0$ in the case of the three-layer model.}
\label{figfin}
\end{center}
\end{figure}

That this is so, can be easily verified by comparing the instantaneous dissipation time with
the nonlinear time $t_{NL} = 1/(ku)$ (with $k=k_\bot$ being the largest forcing scale, and $u$ the rms velocity) which usually rules the direct Kolmogorov cascade.
However, a strikingly good result (shown in Fig.~\ref{figfin} as a red line, while the standard nonlinear time
is shown in blue) is obtained when using a factor $V_a^c/u$ to increase the nonlinear time $1/(ku)$:
\begin{equation}
t_D \simeq 1/(ku) (V_a^c/u)
\label{IKIK}
\end{equation}

The fit is good enough and works for all values listed in Table~\ref{table1} for the turbulence strength $\chi_0$ for the 3-layer model. However, the fit is more hazardous for the 1-layer model, and simply does not work at all for the closed model.

This simple law for the three-layer model deserves some comments.
Note that the expression in Eq.~\ref{IKIK} reminds one of the delayed cascade time predicted by the Iroshnikov-Kraichnan phenomenology.
This phenomenology was meant to describe the (delayed) cascade of interacting Alfv\'en waves with wavevectors not perpendicular to the mean field. The delaying effect was supposed to work all along the cascade, leading to a specific spectrum, different from the Kolmogorov one (3/2 instead of 5/3).
Here, the situation is different. Indeed, the coronal spectrum adopts a slope close to 5/3, not 3/2, and,
most probably (but this study is postponed to a later work), the characteristic time scale which rules the cascade here is the strong turbulence time scale $1/(ku)$.
However, the energy is dominated by the spectral hump at large scale which is not ruled by the fast time scale, but by the delayed time scale Eq.~\ref{IKIK}.
This comes from the effect of the resonant trapped linear modes which act to deplete the nonlinear coupling terms, in the same way as in the Iroshnikov-Kraichnan phenomenology, although only at large scales.

\subsection{Conclusion}

We focus here on the 3-layer model which is by far the most realistic model of the three we have studied.
We have increased our knowledge concerning the physical mechanisms at work since we know now that 
a) the coupling between the chromospheric input and
that of the coronal cavity is close to that of zero-frequency resonance 
b) leakage always plays a substantial role 
c) the time scale of dissipation is long, being the Kolmogorov time $1/(ku)$ reduced by the $u/V_a^c$ factor.

We also have found that a rough prediction for the dissipation rate is given by the classical expression:
$D \simeq D_H = z_{TR}^2/t_a^c$ (Eq.~\ref{hol}, see Fig.\ref{triple-bis})
where $z_{TR}$ is the amplitude of the input fluctuation at the T.R. level.
This result is a bit paradoxical, since this relation has been first obtained by Hollweg, on the basis of assumptions which are not verified in our simulations: (i) a short correlation time for the chromospheric input (ii) negligible leakage.
In our simulations the conditions are completely different: (i) long correlation time (ii) substantial leakage.
The solution of the paradox lies in the rough compensation of several effects:
(i) the energy level is decreased due to leakage and to the chromospheric
turbulence, but largely increased due to resonance 
(ii) the dissipation time is increased due to the large-scale hump.

Although the above heating rate contains the unknown T.R. level of input
fluctuations, it can be used as a predictive law if we identify the T.R. input
value $z_{TR}$ with the (imposed) photospheric value $U_0$. This gives the simple classical result
\begin{equation}
D = D_{H0} = U_0^2/t_a^c
\label{hol0}
\end{equation}
If we consider again the 3-layer results and plot the ratio $D/D_{H0}$ instead of 
the ratio $D/D_{H}$ as in Fig.~\ref{triple-bis}, it is interesting to note that the result is not basically changed, but nevertheless the deviation from 
the horizontal (here $D/D_{H0}=1$) is a bit reduced, being limited to at most a factor $5$.

Can we use our new knowledge to improve the prediction of the dissipation rate beyond the approximate law $D \simeq D_H$ ?
Unfortunately the answer is no, because of our poor knowledge concerning the relation between the known photospheric input $U_0$ and the (largely unknown) chromospheric input $z_{TR}$ as well as the coronal velocity fluctuation level $u_c$.
If we bypass this step by replacing the unknown quantities ($z_{TR}, u_c$) by $U_0$, using the zero-frequency resonant expressions with dominant leakage for the coronal fluctuations 
($b_c=U_0/\epsilon$, $u_c = U_0$, Eq.~\ref{chi3}) 
and the dissipation time (Eq.~\ref{IKIK}), 
we obtain for the dissipation rate per unit mass:
\begin{eqnarray}
D &=& b_c^2/t_D \simeq U_0^2/\epsilon^2 (1/t_{NL}^0) (U_0/V_a^c)\nonumber\\
&=& D_H \ (L/l_\bot) \ (U_0/V_a^0)^2
\label{SUPER}
\end{eqnarray}
When using this expression with the parameter values of the 3-layer model as
given in Table~\ref{table1}, one obtains the dashed line in fig~\ref{triple-bis}. 
This is clearly not an improvement of the simple relation $D/D_H \simeq 1$.
It is actually much worse, by direct comparison with the numerical simulation results
(the circles in fig~\ref{triple-bis}) but also from a more general point of view, 
since dissipation is largely believed to grow with $B_0$, while Eq.~\ref{SUPER} predicts the reverse ($D \propto 1/V_a^0$). 

To progress, we must understand how to relate the chromospheric and coronal velocity level to the photospheric one.
We should also explore how the heating rate depends on all parameters, in
particular the Alfv\'en speed contrast $\epsilon$ and the loop length $L_c$.
Finally, we should investigate if the properties of photospheric turbulence, in particular the correlation time, modifies or not the coronal reaction.

%%%%%%%%%%%%%%%%%%%%% FIN

\begin{acknowledgements}
We benefited from useful discussions with G. Belmont. 
A.V. acknowledges support from the Belgian Federal Science Policy Office
through the ESA-PRODEX program.
The research described in this paper was carried out in part at the Jet
Propulsion Laboratory, California Institute of Technology, under a contract
with the National Aeronautics and Space Administration.
\end{acknowledgements}

\begin{appendix}
\section{Atmospheric model}\label{appendix1}
In order to obtain the parameter $\chi_0$ for "realistic" coronal loop, we
have to determine the Alfv\'en crossing time, or in other words the relation
$\epsilon(L)$ for a given $V_a^0$. 
We model the loop as a semicircular cylinder 
of radius $R$, subject to a constant vertical acceleration
$g=GM_\odot/R_\odot^2$. The loop has constant cross-section and is threated by a uniform magnetic field. 
For simplicity the loop is assumed to be isothermal in
the chromosphere and in the corona, the two temperatures are related by the
jump at the transition region (T.R)
\begin{eqnarray}
T(s)=T_0+\frac{1}{2}\left(T_c-T_0\right)
\left[\tanh\left(\frac{s-s_{tr}^R}{\delta_{tr}}\right) +
      \tanh\left(\frac{s_{tr}^L-s}{\delta_{tr}}\right)\right]
%\left[\tanh\frac{L/2+s-s_{tr}}{\delta_{tr}} +
%      \tanh\frac{L/2-s-s_{tr}}{\delta_{tr}}\right]
\label{eq:temp}
\end{eqnarray}
where $s\in[0,\pi R]$ is the coordinate along the loop,
$s_{tr}^{L,R}$ and $\delta_{tr}$ are the position and width of the
two transition regions,
$s_{tr}^L= R\sin^{-1}(h_{tr}/R)$, $s_{tr}^R= L-s_{tr}^L$.
We set the T.R. height
$h_{tr}=2~\mathrm{Mm}$ and its width to 
$\delta_{tr}=0.2~\mathrm{Mm}$. We will consider two coronal
temperatures, $T_c=0.8~\mathrm{MK},~3~\mathrm{MK}$ in order to consider short loop
(reaching the low corona) and longer loops.
We finally assign to the "base" parameters, magnetic field, number density, and
temperature, the following values:
$B_0=100~\mathrm{G}$, $n_0=10^{17}~\mathrm{cm^{-3}}$ and $T_0=4500~\mathrm{K}$).\\
The density profile along the loop is obtained by solving the equation
for the static equilibrium
\begin{eqnarray}
\frac{1}{\rho}\ds{\rho}=-\frac{1}{T}\ds{T} - \frac{g}{T}\cos{\pi s/L} 
\label{eq:hydro}
\end{eqnarray}
and the $\cos$ function accounts for the projection of gravity along the loop.
Varying the loop length we will find \textsl{short} loops that don't reach the
T.R. heights ($R<2~\mathrm{Mm}$) and \textsl{long} loops, that indeed
reach the corona. For the former the density is found by direct integration of
the above equation, while for long loops the equations will be solved
numerically.\\ 
The static loop model, according to its temperature profile, 
defines a relation $\epsilon(L)$ that can be estimated by considering the temperature 
jump at the T.R. as a discontinuity and calculating the density at
the loop apex $s=L/2$.
By integrating from the photosphere to the T.R. and from
the T.R. to the corona one finds
\begin{eqnarray}
\ln(\rho_{tr-}/\rho_0)&=&-gh_{tr}/T_0\\
\ln(\rho_c/\rho_{tr+})&=&-g(R-h_{tr})/T_c
\end{eqnarray}
where $\rho_{tr\mp}$ are the densities just below and above the transition
region. Assuming that the T.R. is in pressure equilibrium, the density jump is given by
$\ln(\rho_{tr+}/\rho_{tr-})=\ln(T_0/T_c)$ 
so finally one gets:
\begin{eqnarray}
%\ln(\rho_c/\rho_0)= -Rg/T_c (1-h_{tr}/R) + \ln(T_0/T_c) - g h_{tr}/T_0=\\
2\ln{\epsilon} = - \left[\frac{R-h_{tr}}{H_c}+\frac{h_{tr}}{H_0}\right]
+\ln\left[\frac{T_0}{T_c}\right]
\approx- \left[\frac{h_{tr}}{H_0}\right]
-\ln\left[\frac{T_c}{T_0}\right]
\label{eq:h}
\end{eqnarray}
where we have introduced the density scale heights in the chromosphere and
corona, 
%$H_{0}=T_{0}/g\approx 0.28~\mathrm{Mm}$ and $H_{c}=T_{c}/g\approx
%70~\mathrm{Mm}$ respectively, and make use of the definition
$H_{0}=T_{0}/g\approx 0.27~\mathrm{Mm}$ and $H_{c}=T_{c}/g\approx
60~\mathrm{Mm}$ respectively, and make use of the definition
$\epsilon=(\rho_c/\rho_0)^2$. 
From Eq.~\ref{eq:h} one can see that for long loops, $R<H_c$ (used in
the last equality), the density contrast is determined almost entirely by the T.R. jump and is \textsl{independent} of the loop length, except
for very long loops that span a density scale height in
the corona ($L\gg200~\mathrm{Mm}$).
By numerical integration of Eqs.~\ref{eq:temp}-\ref{eq:hydro} we obtain for a
long loop $\epsilon=0.004$, which is a factor two larger than the estimate
based on Eq.~\ref{eq:h}, the discrepancy arising from the fact that the
T.R. is not in pressure equilibrium.\\
\begin{figure}
\begin{center}
\includegraphics [width=\linewidth]{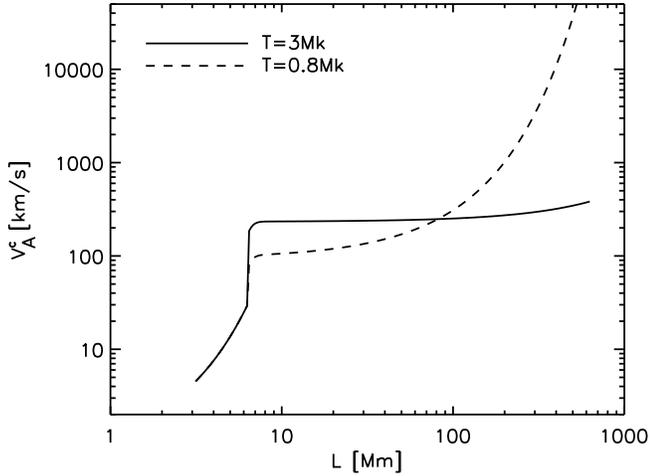}
\caption
{Coronal Alfv\'en speed $V_a^c$ as a function of the loop length $L$ for the
hydrostatic-two temperature model of coronal loop. Solid and dashed line
correspond to coronal temperature of $3$~MK and $0.8$~MK respectively}
\label{hydro}
\end{center}
\end{figure}

To obtain the solid and dashed black lines in Fig.~\ref{fig1a} 
we use the relation $\epsilon(L)=V_a^0/V_a^c(L)$ as found from the
numerical integration for the two coronal temperatures 
$T_c=0.8~\mathrm{MK}$ and $T_c=3~\mathrm{MK}$ (here
$V_a^c(L)\equiv\mathrm{max_x}[V_a(L,x)]$).
The maximal Alfv\'en speed is shown in fig~\ref{hydro} as a function of the
loop length: $V_a^c$ increases monotonically and then experiences a sudden jump
at around $L=4.2$~Mm. After that jump it decreases slightly and then increases
again monotonically. In the first part ($L\lesssim 4.2$~Mm) loop are short enough to remain in the first isothermal layer (the chromosphere), where the density scale height is small. The jump at $L\approx 4.2$~Mm is determined by the fact that the loop reaches the height of the T.R.. In this thin layer the
density scale height is very small, and density drop very quickly. Loops with
length between $\approx4.2$~Mm and $4.4$~Mm don't penetrate into the corona, remaining in the T.R., thus the Alfv\'en speed increases even more, reaching a local maximum. The next part of the
profile is characteristic of loops that reach the second isothermal layer (the
corona), where the density scale height is large. 

%Then we assign the nonlinear timescale by assuming $U_0=1~\mathrm{km/s}$ and
%$l_\bot=6~\mathrm{Mm}$, which gives $t_{NL}^0=l_bot/2\pi U_0\approx 1000~\mathrm{s}$.\\

\end{appendix}
\bibliographystyle{aa} % style aa.bst

\end{document}